\documentclass[12pt]{iopart}
\usepackage{tabularx}
\usepackage{graphicx}
\usepackage{bm}
\usepackage{exscale}
\usepackage{amssymb}
\usepackage{amsfonts}
\usepackage{amsbsy}

\def\text#1{\mbox{\rm #1}}
\def\eqref#1{\ref{#1}}

\def\vecQ{{\bf Q}}

\def\be{\begin{equation}}
\def\ee{\end{equation}}
\def\ba{\begin{eqnarray}}
\def\ea{\end{eqnarray}}

\def\LSCO{La$_{2-x}$Sr$_x$CuO$_4$}
\def\LBCO{La$_{2-x}$Ba$_x$CuO$_4$}
\def\YBCO{YBa$_2$Cu$_3$O$_{6+x}$}
\def\HBCO{HgBa$_2$Cu$_3$O$_{4+\delta}$}

\def\C60{A$_x$C$_{60}$}
\def\LNSCO{La$_{1.6-x}$Nd$_{0.4}$Sr$_x$CuO$_{4}$}

\def\LCO{La$_2$CuO$_4$}

\def\BSCCO{Bi$_2$Sr$_2$CaCu$_2$O$_{8+\delta}$}

\def\LNSCO{La$_{1.6-x}$Nd$_{0.4}$Sr$_x$CuO$_{4}$}

\def\HgCu3{HgCa$_2$Cu$_3$O$_{8+y}$}
\def\HgCu4{HgBa$_2$Ca$_3$Cu$_4$O$_{10+y}$}
\def\TlCu{Tl$_2$Ba$_2$CuO$_{6+\delta}$}
\def\TlCu3{Tl$_2$Ba$_2$Ca$_2$Cu$_3$O$_{10+y}$}
\def\TlCu4{Tl$_2$Ba$_2$Ca$_3$Cu$_4$O$_{12+y}$}

\def\BiCu3{Bi$_2$Sr$_2$Ca$_{2}$Cu$_3$O$_y$}

\def\8LSCO{La$_{1.88}$Sr$_{.12}$CuO$_4$}
\def\110LNSCO{La$_{1.5}$Nd$_{0.4}$Sr$_{0.1}$CuO$_{4}$}
\def\stage4LCO{La$_{2}$CuO$_{4+\delta}$}
\def\Y248{YBa$_2$Cu$_4$O$_8$}

\def\NbSe2{NbSe$_2$}
\def\TaSe2{TaSe$_2$}
\def\TiSe2{TiSe$_2$}
\def\NaCoOH2O{Na$_{0.3}$CoO$_{2y}$H$_2$O}
\def\MgB2{MgB${}_2$}

\def\s12{spin-${\frac{1}{2}}$}

\begin{document}

\title[Striped superconductors]{Striped superconductors: How the cuprates intertwine spin, charge and
superconducting orders}
\author{Erez Berg,$^1$ Eduardo Fradkin,$^2$ Steven A. Kivelson,$^1$ and John
M. Tranquada$^3$}

\address{$^1$Department of Physics, Stanford University, Stanford,
California 94305-4060, USA}
\address{$^2$Department of Physics, University
of Illinois at Urbana-Champaign, Urbana, Illinois 61801-3080, USA}
\address{$^3$Condensed Matter Physics \&\ Materials Science Department,
Brookhaven National Laboratory, Upton, New York 11973-5000, USA}

\date{\today }

\begin{abstract}
Recent transport experiments in the original cuprate high temperature
superconductor, {La$_{2-x}$Ba$_x$CuO$_4$}, have revealed a remarkable
sequence of transitions and crossovers which give rise to a form of
dynamical dimensional reduction, in which a bulk crystal becomes essentially
superconducting in two directions while it remains poorly metallic in the
third. We identify these phenomena as arising from a distinct new
superconducting state, the ``striped superconductor,'' in which the
superconducting order is spatially modulated, so that its volume average
value is zero. Here, in addition to outlining the salient experimental
findings, we sketch the order parameter theory of the state, stressing some
of the ways in which a striped superconductor differs fundamentally from an
ordinary (uniform) superconductor, especially concerning its response to
quenched randomness. We also present the results of DMRG calculations on a
model of interacting electrons in which sign oscillations of the
superconducting order are established. Finally, we speculate concerning the
relevance of this state to experiments in other cuprates, including recent
optical studies of {La$_{2-x}$Sr$_x$CuO$_4$} in a magnetic field, neutron
scattering experiments in underdoped {YBa$_2$Cu$_3$O$_{6+x}$}, and a host of
anomalies seen in STM and ARPES studies of {Bi$_2$Sr$_2$CaCu$_2$O$_{8+\delta} $}.
\end{abstract}

\maketitle

\section{Introduction}

In this paper  we 
carefully characterize in terms of its broken symmetries, a novel
superconducting state of matter, the ``pair-density-wave'' (PDW), with
special focus on the ``striped superconductor'', a unidirectional PDW. We
present a concrete microscopic model of interacting electrons which we show,
using density matrix renormalization group (DMRG) methods, has a striped
superconducting ground state. There is an intimate relation between PDW and
charge density wave (CDW) order, as a consequence of
which the striped superconductor exhibits an extreme sensitivity to quenched
disorder which inevitably leads to glassy behavior. This 
is qualitatively different from the familiar effects of disorder in 
uniform
superconductors. 

On the experimental side, we first draw attention to a set of recently
discovered transport anomalies in the high temperature superconductor, {\LBCO}, 
which are particularly prominent for $x=1/8$. We will
be particularly interested in the spectacular dynamical layer decoupling
effects recently observed in this system \cite{li-2007} which 
indicate that the effective inter-layer Josephson coupling
becomes vanishingly small with decreasing temperature. These experiments
suggest that a special symmetry of the state is required to explain this
previously unsuspected behavior. While no comprehensive theory of these
observations currently exists, even at the phenomenological level, we show
how the salient features of these observations can be straightforwardly
rationalized under the assumption that {La$_{2-x}$Ba$_x$CuO$_4$} is a
striped superconductor. We outline some further experiments that could
critically test this assumption. Finally, we speculate about the possible
role of striped superconducting order as the source of a number of salient
experimental anomalies in a much broader spectrum of high temperature
superconductors, including recent experiments on magnetic field induced
layer decoupling in {La$_{2-x}$Sr$_x$CuO$_4$} \cite{basov08}, the notable
evidence of a local gap with many characteristics of a superconducting gap
in STM and ARPES experiments in {Bi$_2$Sr$_2$CaCu$_2$O$_{8+\delta}$} 
\cite{howald-2003a,lang-2002,kohsaka-2007}, and, most speculatively of all,
experiments indicative of time reversal symmetry breaking in the pseudo-gap
regime of {YBa$_2$Cu$_3$O$_{6+x}$}.

The striped superconductor is a novel state of strongly correlated
electronic matter in which the superconducting, charge and spin order
parameters are closely intertwined with each other, rather than merely
coexisting or competing. As we show here (and discussed in \cite{berg-2007,berg-2008a}) 
the striped superconductor arises from the competing
tendencies existing in a strongly correlated system, resulting in an
inhomogeneous state in which all three forms of order are simultaneously
present. The striped superconductor 
is thus a new type of electronic liquid crystal state \cite{kivelson-1998}.\footnote{Electronic liquid crystals\cite{kivelson-1998} are quantum states of matter that 
spontaneously break some of the translation and/or rotational symmetries of an electronic system. 
In practice, these symmetries are typically
not the continuous symmetries of free space, but rather the various discrete symmetries of the host crystal.
Although an electronic smectic (stripe ordered state) has the same order parameter as a CDW or an SDW, it is a more general state that 
does not
necessarily derive from a nesting vector of an underlying Fermi surface.  The liquid crystal picture offers a broader perspective on the individual phases and on
 their phase transitions\cite{Sun2008}. In particular, the way in which the  PDW phase 
 intertwines charge, spin, and 
 superconducting orders is unnatural in terms of a Fermi surface instability, but not so from the liquid crystalline perspective.}
In particular, as we shall see, the symmetry breaking pattern of the striped
superconductor naturally explains the spectacular layer decoupling effects
observed in experiments in {La$_{2-x}$Ba$_x$CuO$_4$}. In contrast,
in any state with uniform superconducting order, dynamical inter-layer
decoupling could only arise if a somewhat unnatural fine tuning of the
inter-layer couplings led to a sliding phase \cite{ohern-1999}.

The striped superconductor has an order parameter describing a paired state
with non-vanishing wave vector, ${\bf Q}$, the ordering wave vector of the
unidirectional PDW. As such, this state is closely related by symmetry to
the Fulde-Ferrell \cite{Fulde-1964} (FF) and Larkin-Ovchinnikov \cite{Larkin-1964} (LO) states.
The order parameter structure of the PDW state, involving several order
parameters coupled to each other, also evokes the $SO(5)$ approach of a
unified description of antiferromagnetism and a uniform $d$-wave
superconductor \cite{Zhang-1997,demler04}.
The relation of the present discussion to these other systems and to earlier
theoretical works on the same and closely related subjects is deferred to
Sec. \ref{sec:history}. The physics of stripe phases in the cuprate superconductors has been reviewed in 
Ref.\cite{kivelson-2003} and more recently in Ref.\cite{vojta-2009}.

It is important to stress that the 
macroscopically superconducting phase of the cuprates 
reflects the existence of a spatially uniform ${\bf Q}={\bf 0}$ component of the order parameter, whether or not there is substantial finite range superconducting order at non-zero ${\bf Q}$.  One might therefore reasonably ask whether striped superconductivity, even if interesting in its own right,
is anything but an exotic oddity, with little or no relation to the essential physics of high temperature superconductivity.
The answer to this question is at present unclear, and will {\it not} be addressed to any great extent in the present paper. However, we wish to briefly speculate on a way in which the striped superconductor could play an essential role in the 
broader features of this problem. In BCS theory, the superconducting state emerges from a Fermi liquid in which the strong electronic interactions are already accounted for in the self-energy of the quasiparticles.  The cuprates are different, in that the superconductivity, especially in underdoped materials,  emerges from a pseudogap phase for which there is no commonly accepted model.  As we will show, striped superconductivity has features in common with the pseudogap phase, such as a gapless nodal arc and antinodal gaps.  We speculate that the pseudogap phase might be associated with fluctuating striped superconductivity, a state that we do not yet know how to treat. Nevertheless, analysis of the ordered PDW state and comparison to observations of stripe-ordered cuprates is a starting point.  Indeed, comparison (see Subsection \ref{subsec:lbco}) 
of recent photoemission results on LBCO x=1/8 with transport and optical properties suggests that a ``uniform'' d-wave state ({\it i.e.} one with a non-zero uniform component of the order parameter) develops on top of a striped superconductor, resulting in a fully superconducting Meissner state, albeit one with substantial coexisting short-range correlated stripe order.

The rest of this paper is organized as follows: In Section \ref{sec:order}
we give an order parameter description of the pair-density wave state. In
Secs. \ref{sec:experiment} and \ref{sec:other}, we summarize the
experimental evidence for this state, with Sec. \ref{sec:experiment}
focussing on the strongest case, {La$_{2-x}$Ba$_x$CuO$_4$}, and 
Sec. \ref{sec:other} on other cuprates. In Section \ref{sec:microscopic} we
discuss the microscopic mechanisms for the formation of a PDW state. In
Subsection \ref{sec:models} we implement these microscopic considerations by
constructing a specific model that exhibits a PDW phase. The central
conceptual ingredient is a microscopic mechanism leading to the formation of
$\pi$ junctions in an unidirectional PDW state, which is given in Subsection 
\ref{sec:perturbative} using perturbative arguments and then checked
numerically using the density matrix renormalization group (DMRG) (in
Subsection \ref{sec:DMRG}). A solvable microscopic model is discussed in
Subsection \ref{sec:model-striped_sc}. The quasi-particle spectrum of the
PDW state is discussed in Subsection \ref{sec:qp-spectrum}. Next, the
Landau-Ginzburg theory of the PDW phase is discussed in Section \ref{sec:op}. In Section \ref{sec:Tbreaking} we show that the PDW state, in three
dimensional layered structures (orthorhombic and LTT) as well as at grain
(twin) boundaries, leads to time-reversal symmetry breaking effects. In
Section \ref{sec:history} we discuss the connections that exist between the
PDW state and other states discussed in the literature, particularly the
Fulde-Ferrell-Larkin-Ovchinnikov (FFLO) states. Section \ref{sec:final} is
devoted to our conclusions.

This paper is partly a review of our recent work on the theory of the PDW state\cite{berg-2007,berg-2008a} and of other related work, with an updated discussion of the current experimental status. However in this paper we have also included many new results, particularly the DMRG analysis of PDW states in strongly correlated systems of Section \ref{sec:DMRG}, and the connection bewteen the PDW state and non-collinear order and time-reversal symmetry breaking of Section \ref{sec:Tbreaking}.

\section{The Order Parameter of a Striped Superconductor}
\setcounter{footnote}{0}

\label{sec:order}

The order parameter whose non-zero expectation value defines a
superconducting state is
\begin{equation}
\phi_{\sigma,\sigma^\prime}(\mathbf{r},\mathbf{r}^{\prime}) \equiv \langle
\psi_{\sigma}^\dagger(\mathbf{r})\psi_{\sigma^\prime}^\dagger (\mathbf{r}^\prime)\rangle,
\end{equation}
where $\psi^\dagger_{\sigma}(\mathbf{r})$ is the fermionic field operator
which creates an electron with spin polarization $\sigma$ at position $\mathbf{r}$. Further distinctions between different superconducting states
can be drawn on the basis of the spatial and spin symmetries of $\phi$. In
crystalline solids, all familiar superconducting states respect the
translational symmetry of the solid, $\phi(\mathbf{r}+\mathbf{R},\mathbf{r}^\prime+\mathbf{R}) = \phi(\mathbf{r},\mathbf{r}^\prime)$, 
where $\mathbf{R}$
is any Bravais lattice vector. Consequently, the symmetries of the state can
be classified by the irreducible representations of the point group -
colloquially as s-wave, d-wave, p-wave, etc. In the absence of spin-orbit
coupling, superconducting states can be classified, as well, by their
transformation under spin rotations as singlet or triplet. Finally, the
superconducting state can either preserve or break time reversal symmetry
(as in $p_x+ip_y$).

In the presence of quenched disorder, the underlying Hamiltonian does not
have any particular spatial symmetries, so the classification of distinct
superconducting states by their symmetries (other than time reversal), at
first seems difficult. However, there are several ways that this can be
accomplished \cite{spivak-2008}, of which the most obvious is to consider
the symmetries of the configuration averaged order parameter
\begin{equation}
\overline{ \phi}_{\sigma,\sigma^\prime}(\mathbf{r},\mathbf{r}^\prime) \equiv
\overline{ \langle \psi_{\sigma}^\dagger(\mathbf{r})
\psi_{\sigma^\prime}^\dagger(\mathbf{r}^\prime)\rangle},
\end{equation}
where $\langle ... \rangle$ signifies the thermal average, and $\overline {(
\ldots )}$ signifies an average over realizations of the disorder
configuration. It is clear, for example, that under most circumstances, a
macroscopic ``phase sensitive'' measurement of the symmetry of the order
parameter will give \cite{spivak-2008} a result consistent with a
classification based on the symmetry of the configuration averaged order
parameter.

The striped superconductor is an example of a state, which has more
generally called \cite{chen-2004,tesanovic-2005,balents05a} a pair density
wave (PDW), in which the translational symmetry of the crystal is
spontaneously broken as well, so that 
$\phi(\mathbf{r}+\mathbf{R}, \mathbf{r}^\prime+\mathbf{R})$ exhibits non-trivial dependence on $\mathbf{R}$.
However, this by itself, is insufficient to identify a new state of matter.
In a system with coexisting charge-density-wave (CDW) and superconducting
order, the CDW itself introduces a new periodicity into the problem, which
must generically be reflected in a spatial modulation of $\phi $, as well.\footnote{The problem of coexisting stripe and superconducting order in strongly correlated systems has been the focus of numerous studies in the literature. Sachdev, Vojta, and coworkers have investigated this problem in detail in the context of generalized 2D $t-J$ models in the large $N$ approximation\cite{vojta-1999,vojta-2000,vojta-2008}. This problem has also been discussed in one-dimensional systems\cite{Aligia-2007}.}
As discussed in Sec. \ref{sec:op}, an analysis of the implications of a
generic theory of coupled order parameters implies \cite{berg-2008a} that in a
state of coexisting order, a (possibly small) modulation of the
superconducting order with the same spatial period as that of the CDW will
be induced. None-the-less, in such a state, there still exists a
``dominant'' uniform component to the superconducting order parameter, which
we define as the spatial average of the SC order parameter:
\begin{equation}
\phi^{(0)}_{\sigma,\sigma^\prime}(\mathbf{r}, \mathbf{r}^\prime) \equiv
N^{-1}\sum_{\mathbf{R}} \langle \psi_{\sigma}^\dagger(\mathbf{r}+\mathbf{R})
\psi_{\sigma^\prime}^\dagger(\mathbf{r}^\prime+\mathbf{R})\rangle,
\end{equation}
where $N$ is the number of unit cells in the system.

Instead, the pure PDW in a crystal is a state in which $\phi$ is non-zero,
but all uniform components vanish, 
$\phi^{(0)}_{\sigma,\sigma^\prime}(\mathbf{r}, \mathbf{r}^\prime)=0$ for any $\mathbf{r}$ and $\mathbf{r}^\prime $. 
Just as a CDW is often defined in terms of a fundamental
harmonic, so a PDW state is characterized by the smallest value of the
crystal momentum, $\mathbf{Q}$, for which
\begin{equation}
\phi^{(\mathbf{Q})}_{\sigma,\sigma^\prime}(\mathbf{r}, \mathbf{r}^\prime)
\equiv N^{-1}\sum_{\mathbf{R}} \exp[i\mathbf{Q} \cdot \mathbf{R}] \langle
\psi_{\sigma}^\dagger(\mathbf{r}+\mathbf{R}) \psi_{\sigma^\prime}^\dagger(\mathbf{r}^\prime+\mathbf{R})\rangle,
\end{equation}
has a non-vanishing expectation value.\footnote{
As with a uniform superconducting state, distinct PDW states with the same
pattern of translation symmetry breaking can also be distinguished by
different patterns of point group symmetry breaking. However, since the
ordering vector (or vectors) already break the point group down to a smaller
subgroup, which is then all that is left of the original symmetry for this
purpose. For instance, in a tetragonal crystal, a striped superconductor
with $\mathbf{Q}$ along the x direction, can be classified as having 
$s$-wave or $d_{xy}$-wave symmetry, based on whether or not the the order
parameter changes sign under reflection through a symmetry plane parallel to
the x axis, but any distinction one would like to draw between a striped
version of an s-wave and a d$_{x^2-y^2}$-wave superconductor are in precise,
not based on broken symmetries but on quantitative differences in local
pairing correlations, and so do not define distinct phases of matter.
However, a checkerboard PDW, with symmetry related ordering vectors 
$\mathbf{Q}$ and $\mathbf{Q}^\prime$ along the x and y axes, respectively, can be
classified as s-wave or $d_{x^2-y^2}$-wave, depending on how it transforms
under rotation by $\pi/2$.}

Note that the theory of coupled order parameters \cite{berg-2008a} implies that
the existence of PDW order with ordering vector $\mathbf{Q}$ generically
implies the existence of CDW order with ordering vector $2\mathbf{Q}$, but
so long as $\phi^{0}=0$, no CDW ordering with wave vector $\mathbf{Q}$ is
expected. A ``striped superconductor'' refers to the special case in which
the independent ordering vectors are all parallel to each other
(``unidirectional PDW'').

One of the prime new characteristics of a striped superconductor which is
different from a uniform superconductor is its complex sensitivity to
quenched disorder. As we shall see, for much the same reasons that disorder
destroys long-range CDW order, under most relevant circumstances, even weak
disorder causes the configuration averaged PDW order parameter to vanish:
\begin{equation}
\overline{\phi}_{\sigma,\sigma^\prime}(\mathbf{r}, \mathbf{r}^\prime) =0.
\end{equation}
However, as in the case of an XY spin-glass, this is not the whole story: It
is possible to define an analogue of the Edwards-Anderson order parameter,
\begin{equation}
Q_{\sigma,\sigma^\prime}(\mathbf{r},\mathbf{r}^\prime)\equiv \overline{
|\langle \psi_{\sigma}^\dagger(\mathbf{r}) \psi_{\sigma^\prime}^\dagger(\mathbf{r}^{\,\prime})\rangle |^2},
\end{equation}
which vanishes in the normal high temperature phase, but which can be
non-zero in a low temperature superconducting glass phase, where one exists.
Moreover, in such a phase, as we will see, we generically expect
time-reversal symmetry to be spontaneously broken. In analogy with the XY
spin-glass, we expect that in two dimensions, the superconducting glass
phase is stable only at $T=0$ and for weak enough disorder, although in
three dimensions it can exist below a non-zero superconducting glass
transition temperature.\footnote{
In \cite{berg-2008a}, the possibility is discussed that in three dimensions
there might also exist a superconducting version of a Bragg glass phase, in
which $\phi$ exhibits quasi-long-range order. We have not further studied
this potentially interesting state.}

There is one more extension that is useful---we define a charge $4e$
superconducting order parameter:
\begin{eqnarray}
\phi^{(4)}(1,2,3,4) \equiv \langle \psi_{\sigma_1}^\dagger(\mathbf{r}_1)\psi_{\sigma_2}^\dagger(\mathbf{r}_2) \psi_{\sigma_3}^\dagger(\mathbf{r}_3)\psi_{\sigma_4}^\dagger(\mathbf{r}_4)\rangle  \label{phi4}
\end{eqnarray}
where we have introduced a compact notation in which $1 \equiv (\sigma_1,\mathbf{r}_1)$, etc. 
Naturally, in any state with charge 2e superconducting
order, $\phi_{\sigma,\sigma^\prime}(\mathbf{r},\mathbf{r}^\prime)\neq 0$,
some components of the charge 4e order parameter will also be non-zero. This
can be seen from the theory of coupled order parameters presented in Sec. 
\ref{sec:op}. At mean-field level, it can be seen by applying Wick's theorem
to the expression in Eq.~\ref{phi4} to express $\phi^{(4)}$ as a sum of
pairwise products of $\phi$'s: $\phi^{(4)}(1,2,3,4)\sim \phi(12)\phi(34) +
\phi(14)\phi(23)-\phi(13)\phi(24)$.

There are two reasons to consider this order parameter. In the first place,
it is clear from the above that even in the PDW state, although the uniform
component of $\phi$ vanishes, the uniform component of 
$\phi^{4} \sim \phi^{\mathbf{Q}}\phi^{-\mathbf{Q}} \neq 0$. More importantly, $\phi^{(4)}$
ordering can be more robust than the PDW ordering. Specifically, under some
circumstances \cite{berg-2008a,Radzihovsky-2008}, it is possible for thermal
or quantum fluctuations to destroy the PDW order by restoring translational
symmetry without restoring large-scale gauge symmetry; in this case,
appropriate components of $\phi^{(4)}$ remain non-zero, even though $\phi$
vanishes identically. This is the only potentially realistic route we know
of to charge $4e$ superconductivity.\footnote{
It is possible to cook up models in which charge $4e$ superconductivity
arises in systems in which electrons can form 4 particle bound-states, but
do not form 2 particle or many particle bound states (phase separation) -
see, for example, \cite{kivelson-90b}. However, this involves
unrealistically strong attractive interactions and an unpleasant amount of
fine tuning of parameters.}

In the absence of spin-orbit coupling, distinct phases with translationally
invariant charge $4e$ ordering can be classified according to the total spin
of the order parameter, which in this case can be spin 2, 1, or 0.
Manifestly, any charge $4e$ superconducting state which results from the
partial melting of a singlet PDW will itself have spin 0. As with paired
superconductors, the charge 4e order parameter can also be classified
according to its transformation properties under action of the point-group
of the host crystal. For instance, in a crystal with a $C_4$ symmetry,
taking the points $\mathbf{r}_j$ to lie on the vertices of a square, the
transformation properties of $\phi^{(4)}$ under rotation by $\pi/2$ can be
used to classify distinct spin-0 states as being d-wave or s-wave.

The definitions given here in terms of possible behaviors of the order
parameter are natural from a taxonomic viewpoint. In particular, the striped
superconductor seems at first to be a rather straightforward generalization
of familiar uniform superconducting states. However, both at the microscopic
level of the ``mechanism'' of formation of such a state, and at the
phenomenological level of macroscopically observable implications of the
state, the problem is full of subtleties and surprises, as discussed below.

\section{Striped superconductivity in La$_{2-x}$Ba$_x$CuO$_4$ and the 214
family}
\setcounter{footnote}{0}

\label{sec:experiment}

We now summarize some of the observations that lead to the conclusion that La$_{2-x}$Ba$_x$CuO$_4$ with $x=1/8$ is 
currently the most promising candidate
experimental system as a realization of a striped superconductor \cite{li-2007,berg-2007}.

Firstly, the existence of ``stripe order'' is unambiguous. It is well known
(from neutron \cite{fujita-2004,tranquada08,dunsiger-2008} and X-ray \cite{abbamonte-2005,kim-2008,tranquada08} scattering studies) 
that
unidirectional CDW (charge-stripe) and SDW (spin-stripe) orders exist in {La$_{2-x}$Ba$_x$CuO$_4$}.
Such spin and charge stripe orders were originally studied in La$_{1.48}$Nd$_{0.4}$Sr$_{0.12}$CuO$_4$ \cite{tranquada-1995,zimmermann-1998,christensen-2007}, and they have now been
confirmed in La$_{1.8-x}$Eu$_{0.2}$Sr$_x$CuO$_4$ \cite{fink-2008,hucker-2007,teitelbaum-2000}. 
Furthermore, substantial spin
stripe order has been observed in La$_2$CuO$_{4.11}$ \cite{lee-1999},
Zn-doped La$_{2-x}$Sr$_x$CuO$_4$\ \cite{hirota-2001}, and in the spin-glass
regime of La$_{2-x}$Sr$_x$CuO$_4$\ ($0.02 < x <0.055$) \cite{fujita-2002}.
While the spin-stripe order in underdoped but superconducting La$_{2-x}$Sr$_x $CuO$_4$\ ($0.055 < x < 0.14$) is weak in the absence of an applied
magnetic field, it has been observed \cite{lake-2002,khaykovich-2005,chang-2008} that readily accessible magnetic
fields (which partially suppress the superconducting order) produce
well-developed and reproducible spin-stripe order. For {La$_{2-x}$Ba$_x$CuO$_4$} with $x=1/8$, the charge ordering temperature is 53~K and the spin
ordering temperature is 40~K \cite{tranquada08}.

In addition to spin and charge ordering, {La$_{2-x}$Ba$_{x}$CuO$_{4}$} with $x=1/8$ exhibits transport and thermodynamic behavior that is both striking
and complex. We will not rehash all of the details here (see \cite{tranquada08}); however, there are two qualitative features of the data on
which we would like to focus: 1) With the onset of spin-stripe order at 40~K,\footnote{Static charge stripe order onsets at 54 K, at the LTT/LTO transition.\cite{abbamonte-2005}} there is a large (in magnitude) and strongly temperature
dependent enhancement of the anisotropy of the resistivity and other
properties, such that below 40 K the in-plane charge dynamics resemble those
of a superconductor, while in the $c$-direction the system remains poorly
metallic. The most extreme illustration of this occurs in the temperature
range $10\; \mbox{\rm K} < T <16 \; \mbox{\rm K}$, in which the in-plane
resistivity is immeasurably small, while the $c$-axis resistivity is in the
1--10 m$\Omega $-cm range, so that the resistivity anisotropy ratio is
consistent with infinity. 2) Despite the fact that signatures of
superconductivity onset at temperatures in excess of 40 K, and that angle
resolved photoemission (ARPES) has inferred a ``gap'' \cite{valla-2006,shen08} of
order 20 meV, the fully superconducting state (\textit{i.e.}, the Meissner
effect and zero resistance in all directions) only occurs below a critical
temperature of 4 K. It is very difficult to imagine a scenario in which a
strong conventional superconducting order develops locally on such high
scales, but fully orders only at such low temperatures in a system that is
three dimensional, non-granular in structure, and not subjected to an
external magnetic field.

Evidence that similar, although somewhat less extreme transport and
thermodynamic anomalies accompany stripe ordering can be recognized, in
retrospect, in other materials in the 214 family. For example, in {La$_{2-x}$Sr$_x$CuO$_4$} with $x=0.08$ and 0.10, the anisotropy of the resistivity ($c$-axis vs.\ in-plane) rapidly grows towards $10^4$ as the superconducting $T_c $ is approached from above \cite{komiya-2002}. In the case of La$_{1.6-x} $Nd$_{0.4}$Sr$_{x}$CuO$_4$, evidence for dynamical layer decoupling
is provided by measurements of the anisotropic onset of the Meissner effect
\cite{ding-2008}. In contrast, the resistivity ratio in this material \cite{xiang-2008} only reaches $10^3$; this may be limited by enhanced in-plane
resistivity due to disorder \cite{fujita-2005}. Moreover, an unexpectedly
strong layer decoupling in the charge dynamics produced by the
application of a transverse magnetic field in {\LSCO} has
been observed \cite{basov08}, but only in the underdoped range of $x$ where
the magnetic field also induces spin stripe order \cite{lake-2002,chang-2008}.

We shall see that the anomalous sensitivity of a striped superconductor to
quenched disorder can account for the existence of a broad range of
temperatures between the onset of strongly developed superconducting
correlations on intermediate scales and the actual macroscopic transition
temperature to a state of long-range coherence. Moreover, given the crystal
structure of the Low Temperature Tetragonal (LTT) phase of {\LBCO}, there is a special symmetry of the striped superconducting state
which produces interlayer decoupling. Specifically, because the stripes in
alternate planes are oriented perpendicular to one another (as shown in Fig.\ref{fig:spiral}a), there is no first order Josephson coupling between
neighboring planes. Indeed, analogous features of the spin-stripe order,
which have gone largely unnoticed in the past, are accounted for \cite{berg-2007} by the same geometric features of the striped state. When
spin-stripe order occurs, the in-plane correlation length can be very long,
in the range of 100--600~\AA\ \cite{lee-1999,tranquada08}, but the
interplanar correlation length is never more than a few \AA\ \cite{lee-1999,tranquada-1996}, a degree of anisotropy that cannot be reasonably
explained simply on the basis of the anisotropy in the magnitude of the
exchange couplings \cite{sudipprivatecommun}. Furthermore, despite the
presence of long correlation lengths, true long-range spin-stripe order has
never been reported. It will be made clear that the superconducting stripe
order and the spin stripe order share the same periodicity and the same
geometry, so both the interlayer decoupling and the suppression by quenched
disorder of the transition to a long-range ordered state can be understood
as arising from the same considerations applied to both the unidirectional
SDW and PDW orders.

\section{Experiments in other cuprates}
\setcounter{footnote}{0}
\label{sec:other}

Data which clearly reveal the existence of spin-stripe order, or that
provide compelling evidence of of PDW order in other families of cuprates is
less extensive. However, considerable evidence of a tendency to spin stripe
order in {YBa$_2$Cu$_3$O$_{6+x}$} has started to accumulate\cite{kivelson-2003}, 
and there are some persistent puzzles concerning the
interpretation of various experiments in a number of cuprates that, we would
like to speculate, may reflect the presence of PDW order. In this section,
we will mention some of these puzzles, and will return to discuss why they
may be indicative of PDW order in Sec. \ref{sec:Tbreaking}.

{\YBCO}
is often regarded as the most ideal cuprate, having minimal structural and
chemical disorder, and
less tendency to stripe or any other type of charge ordering than the 214
cuprates. (Sometimes the cuprates with the highest transition temperatures,
such as {\HBCO}, are viewed as being similarly
pristine.) However, in its underdoped regime it is well known that {\YBCO} exhibits temperature-dependent in-plane anisotropic transport
\cite{ando-2002} as well as fluctuating spin stripe order \cite{mook-1998,kivelson-2003,stock-2004}. Recent neutron scattering experiments
have provided strong evidence that underdoped {YBa$_2$Cu$_3$O$_{6+x}$} (with
$x\sim 0.45$) has nematic order below a critical temperature $T_c \sim 150$~K \cite{hinkov-2007b}. Even more recent neutron scattering experiments by
Hinkov \textit{et al.} \cite{hinkov-2008} on the same sample find that a
modest $c$-axis magnetic field stabilizes an incommensurate static spin
ordered state, detectable as a pair of peaks in the elastic scattering
displaced by a distance in the crystallographic $a$ direction from the Ne\'{e}l ordering vector. 
Given the newfound evidence of spin-stripe related structures in {\YBCO}, it is plausible that here, too, striped superconductivity may
occur. However, the differences in the 3D crystal structure, and especially
the weak orthorhombicity, would make the macroscopic properties of a PDW
distinctly different in {YBa$_2$Cu$_3$O$_{6+x}$} than in the 214 cuprates.

A remarkable recent discovery is that underdoped {\YBCO}
appears to exhibit signatures of spontaneous time-reversal symmetry breaking
(at zero magnetic field) below a critical temperature comparable to that for
the
nematic ordering. \cite{fauque-2006,xia-2007} ({\HBCO} \cite{greven-2008} exhibits similar
signatures.) Various theoretical scenarios for the existence of
time-reversal symmetry breaking predated these experiments, and so in some
sense predicted them \cite{Varma2005,chakravarty-2001c}. However, given that
\emph{both} nematic order and time-reversal symmetry breaking are seemingly
present simultaneously in the same samples with comparable critical
temperatures, it is reasonable to 
hope that 
both phenomena have an underlying common explanation. If we think of the
superconducting order parameter as an XY pseudo-spin, then the PDW order is
a form of collinear antiferromagnetism, and time-reversal symmetry breaking
corresponds to non-collinear order of the pseudospins. As we will show in
Sec. \ref{sec:Tbreaking}, weak time reversal symmetry breaking can occur in
a PDW state due to various patterns of geometric frustration in three
dimensions or as a consequence of the existence of certain types of defects,
such as twin boundaries. (See, also, \cite{berg-2008a}. )
There is a large body of STM and ARPES data, especially on Bi$_2$Sr$_2$CaCu$_2$O$_{8+\delta}$\ and Bi$_2$Sr$_2$CuO$_{6+\delta}$, which has revealed a
surprisingly rich and difficult to interpret set of spectral features
associated with the d-wave superconducting gap and a d-wave pseudo-gap whose
origin is controversial. Indeed, there is a clear ``nodal-anti nodal
dichotomy'' \cite{zhou04,yang08} in the behavior of the measured
single-particle spectral functions. Some aspects of the data are suggestive
that there is a single superconducting origin of all gap features, with
anisotropic effects of superconducting fluctuations leading to the observed
dichotomy. Other aspects suggest that there are at least two distinct
origins of the near-nodal and the antinodal gaps. It is possible that PDW
ordering tendencies can synthesize both aspects of the interpretation.
In the presence of both uniform and PDW superconducting order, there are two
distinct order parameters, both of which open gaps on portions of the Fermi
surface, 
but they are both superconducting, and so they can smoothly evolve into one
another.
(Note that an early study \cite{podolsky} of modulated structures seen in
STM \cite{hoffman-2002b,howald-2003a,vershinin-2004} concluded that they could be
understood in terms of just such a two-superconducting-gap state.)

More generally, one of the most remarkable features of the pseudo-gap
phenomena is the existence of what appears to be superconducting
fluctuations, detectable \cite{wang-2002b,sri,vadim} for instance in the
Nernst and magnetization signal, over a surprisingly broad range of
temperatures and doping concentrations. At a broad-brush level \cite{emery-1995b}, these phenomena are a consequence of a phase stiffness scale
that is small compared to the pairing scale. However, it is generally
difficult to understand the existence of such a broad fluctuational regime
on the basis of any sensible microscopic considerations. The glassy nature
of the ordering phenomena in a PDW may hold the key to this central paradox
of HTC phenomenology, as it gives rise to an intrinsically broad regime in
which superconducting correlations extend over large, but not infinite
distances.

\section{Microscopic considerations}
\setcounter{footnote}{0}
\label{sec:microscopic}

From a microscopic viewpoint, the notion that a {PDW\ } phase could be
stable at first sounds absurd. Intuitively, the superconducting state can be
thought of as the condensed state of charge $2e$ bosons. However, in the
absence of magnetic fields, the ground-state of a bosonic fluid is always
node-less, independent of the strength of the interactions, and therefore
cannot support a state in which the superconducting order parameter changes
sign. Thus, for a PDW state to arise, microscopic physics at scales less
than or of order the pair-size, $\xi_0$, must be essential. This physics
reflects an essential difference between superfluids of paired fermions and
preformed bosons \cite{spivak-1991}.

Our goal in this section is to shed some light on the mechanism by which
strongly interacting electrons can form a superconducting ground-state with
alternating signs of the order parameter. We will consider the case of a
unidirectional (striped) superconductor, but the same considerations apply
to more general forms of PDW order. We will not discuss the origin of the
pairing which leads to superconductivity. Likewise, we will not focus on the
mechanism of translation symmetry breaking by the density wave, as that is
similar to the physics of CDW and SDW formation. Our focus is on the \emph{sign alternation} of $\phi$. Thus, in much of this discussion, we will adopt
a model in which we have alternating stripes of superconductor and
correlated insulator. The system looks like an array of extended
superconductor-insulator-superconductor (SIS) junctions, and we will
primarily be concerned with computing the Josephson coupling across the
insulating barriers. If the effective Josephson coupling is positive, then a
uniform phase (normal) superconducting state is favored, but if the coupling
is negative (favoring a $\pi$ junction), then a striped superconducting
phase is found.

So long as time reversal symmetry is neither spontaneously nor explicitly
broken, the Josephson coupling, $J$ between two superconductors must be
real. If it is positive, as is the usual case, the energy is minimized by
the state in which the phase difference across the junction is 0; if it is
negative, a phase difference of $\pi$ is preferred, leading to a ``$\pi$
junction.'' $\pi$ junctions have been shown, both theoretically and
experimentally, to occur for two distinct reasons: they can be a consequence
of strong correlation effects in the junction region between two
superconductors \cite{spivak-1991,Rozhkov-1999,vandam-2006} or due to the
internal structure (\textit{e.g.} d-wave symmetry) of the superconductors,
themselves \cite{wollman-1993,tsuei-1994}.

In Ref. \cite{berg-2008a}, we have provided examples of $\pi$ junctions
which build on the first set of ideas. Unlike the previously studied cases,
these $\pi$ junctions were \emph{extended} (i.e., $J$ is proportional to the
cross sectional ``area'' of the junction). However, since the problem was
solved analytically (treating the tunneling between the superconducting and
the insulating regions by perturbation theory), we were limited to somewhat
artificial models. For example, tunneling between the sites in the
insulating regime was neglected. In this Section, we first summarize the
perturbative results of \cite{berg-2008a}, and then present numerical (DMRG)
results for an extended SIS junction. 
Under some circumstances, $J >0$, 
but we also find a considerable region of
parameter space where where $J<0$. Finally, we discuss how this result can
be generalized to an infinite array of junctions, forming a 2D
unidirectional PDW.

\subsection{A Solved model}
\setcounter{footnote}{0}
\label{sec:models}

Let us consider the following explicit model for a single SIS\ junction. The
three decoupled subsystems are described by the Hamiltonian
\begin{equation}
H_{0}=H_{L}+H_{B}+H_{R},  \label{H0}
\end{equation}
The right ($R$) and left ($L$) superconducting regions and the barrier ($B$)
region are one dimensional Hubbard models,
\begin{eqnarray}
H_{\alpha } =\sum_{i\sigma }\left( -tc_{\alpha ,i,\sigma }^{\dagger }
c^{\vphantom{\dagger}}_{\alpha ,i+1,\sigma }+\mathrm{h.c.}-\mu _{\alpha
}n_{\alpha ,i}\right) +U_{\alpha }\sum_{i}n_{\alpha ,i,\uparrow }n_{\alpha
,i,\downarrow\mbox{\rm .} }
\end{eqnarray}
$c_{\alpha ,i+1,\sigma }^{\dagger }$ is a creation operator of an electron
on chain $\alpha =L,R$ or $B$ at site $i$ with spin $\sigma $, and we have
introduced the notation 
$n_{\alpha ,i,\sigma }=c_{\alpha ,i,\sigma}^{\dagger }c_{\alpha ,i,\sigma }^{{}}$ and 
$n_{\alpha ,i}=\sum_{\sigma}n_{\alpha ,i,\sigma }$. The left and right superconducting chains are
characterized by a negative $U_{R}=U_{L}=-\left\vert U_{L,R}\right\vert $,
while the insulating barrier has a positive $U_{B}>0$. The chemical
potentials of the left and right superconductors are the same, $\mu _{R}=\mu
_{L}$, but different from $\mu _{B}$, which is tuned such that the barrier
chain is half filled (and therefore insulating).

The three subsystems are coupled together by a single-particle hopping term,
\begin{equation}
H^{\prime }=-t^{\prime }\sum_{i,\sigma }[c_{L,i,\sigma }^{\dagger }c^{\vphantom{\dagger}}_{B,i,\sigma }+
c_{R,i,\sigma }^{\dagger }c^{\vphantom{\dagger}}_{B,i,\sigma }+\mathrm{h.c.}]\mbox{\rm .}  \label{Hp}
\end{equation}
The left and right chains are characterized by a spin gap and by dominant
superconducting fluctuations, as a result of their negative $U$'s. The
inter-chain hopping term $H^{\prime }$ induces a finite Josephson coupling
between the local superconducting order parameters of the two chains, via
virtual hopping of a Cooper pair through the barrier chain.

\subsubsection{Perturbative analysis}

\label{sec:perturbative}

For completeness, let us briefly review the perturbative treatment of the
inter-chain hopping term (\ref{Hp}) given in \cite{berg-2008a}. The leading
(fourth order) contribution to the Josephson coupling is given by
\begin{equation}
J=\frac{\left( t^{\prime }\right) ^{4}}{\beta }\int \ d1\ d2\ d3\ d4\
F_{L}(1,2)F_{R}^{\star }(4,3)\Gamma (1,2;3,4)  \label{J}
\end{equation}
where $1\equiv (\tau _{1},i_{1})$ etc.,
\begin{equation}
\int d1\equiv \sum_{i_{1}}\int_{0}^{\beta }d\tau _{1}
\end{equation}
(in the limit $\beta \rightarrow \infty $) and
\begin{eqnarray}
F_{\alpha }(1,2) &\equiv &\left\langle T_{\tau }\left[ c_{\alpha
,i_{1},\uparrow }^{\dagger }(\tau _{1})c_{\alpha ,i_{2},\downarrow
}^{\dagger }(\tau _{2})\right] \right\rangle \\
\Gamma (1,2;4,3) &\equiv &\left\langle T_{\tau }\left[ c_{i_{1},\uparrow
}^{\dagger }(\tau _{1})c_{i_{2},\downarrow }^{\dagger }(\tau
_{2})c_{i_{3},\downarrow }(\tau _{3})c_{i_{4},\uparrow }(\tau _{4})\right]
\right\rangle  \nonumber
\end{eqnarray}
where we have made the identification $c_{i,\sigma }^{\dagger }\equiv
c_{B,i,\sigma }^{\dagger }$. Our purpose is to determine the conditions
under which $J<0$.
For the sake of simplicity, let us consider the case in which the gap to
remove a particle from the barrier, $\Delta _{h}$, satisfies $\Delta _{s}\ll
\Delta _{h}\ll \Delta _{p}$, where $\Delta _{s}$ is the spin gap on the
superconducting chains, and $\Delta _{p}$ is the gap to insert a particle in
the barrier. These conditions can be met by tuning appropriately the
chemical potentials on the three chains and setting $U_{B}$ to be
sufficiently large.

In \cite{berg-2008a} it is shown that, quite generally, $J$ can be written
as a sum of two terms 
\begin{equation}
J=J_{1}+J_{2}
\end{equation}
where, in terms of the spin-spin correlation function, $\langle \vec{S}(1)\cdot \vec{S}(2) \rangle$ of the barrier chain,
\begin{eqnarray}
J_{1} &=&\frac{(t^{\prime})^4}{4\beta \left( \Delta _{h}\right) ^{2}}\int \
d1\ d2\left\vert F_{L}(1,2)\right\vert ^{2}  \label{J2} \\
J_{2} &=&-\frac{3(t^{\prime})^4}{4\beta \left( \Delta _{h}\right) ^{2}}\int
\ d1\ d2\left\vert F_{L}(1,2)\right\vert ^{2}\langle \vec{S}(1)\cdot \vec{S}(2) \rangle  \nonumber
\end{eqnarray}
Explicitly, $J_{1}>0$, while for generic circumstances one finds that $J_{2}<0$. The overall sign of $J$ is therefore non-universal, and determined
by which term is bigger. We can, however, identify the conditions under
which $J_{2}$ dominates. Upon a Fourier transform, $\left\vert
F_{L}(1,2)\right\vert ^{2}$ is peaked around two values of the momentum $q$,
at $q=0$ and $2k_{F}$, in which $2k_{F}=\pi n$ where $n$ is the number of
electrons per site in the left and right chains. Since, upon Fourier
transforming, $\langle \vec{S}(1)\cdot \vec{S}(2) \rangle$ is peaked at
momenta $q=0$ and $\pi $, (as can be seen, e.g., from a bosonized treatment
of the half filled chain) we expect that $J_{2}$ in Eq. (\ref{J2}) is
maximized when $n=1$, i.e. when \emph{the superconducting chains are half
filled}. The requirement of proximity to half filling becomes less and less
stringent when $\left\vert U_{R,L}\right\vert $ is increased, since then the
gap $\Delta _{s} $ in the superconducting chains increases and the peaks in $\left\vert F_{L}(q)\right\vert ^{2}$ become more and more broad. These
qualitative expectations are confirmed by numerical DMRG\ simulations,
presented in the next subsection.

\subsubsection{Numerical results}

\label{sec:DMRG}

We have performed DMRG simulations of the model $H=H_{0}+H^{\prime }$ in Eq.
(\ref{H0},\ref{Hp}), with the following parameters: $t=t^{\prime }=1$, $\mu
_{B}=6$, $U_{B}=10$, and variable $U_{L}=U_{R}\equiv -\left\vert
U_{L,R}\right\vert $ and $\mu _{L}=\mu _{R}\equiv \mu _{L,R}$. Most of the
calculations where done with systems of size $3\times 24$. In a small number
of parameter sets, we have verified that the results do not change when we
increase the system size to $3\times 36$. Up to $m=1600$ states where kept
in these calculations. The results (both ground state energies and local
measurements) where extrapolated linearly in the truncation error \cite{white_local}, which is in the range $10^{-5}-10^{-6}$.

In order to measure the sign of the Josephson coupling from the
calculations, we have applied pairing potentials on the left and right
chains, adding the following term to Eq. (\ref{H0}):
\begin{equation}
H_{pair}=-\sum_{i,\alpha =L,R}\Delta _{\alpha }c_{\alpha ,i,\uparrow
}^{\dagger }c_{\alpha ,i,\downarrow }^{\dagger }+\mathrm{h.c.}  \label{Hpair}
\end{equation}
In the presence of this term, the number of particles in the calculation is
conserved only modulo 2. The average particle number is fixed by the overall
chemical potential. Two methods where employed to determine the sign of $J$.
(a) Pairing potentials of either the same sign, $\Delta _{R}=\Delta _{L}$,
and of opposite signs, $\Delta _{R}=-\Delta _{L}$, where applied to the two
chains. The ground state energies in the two cases are $E_{+}$ and $E_{-}$,
respectively. Then $J=E_{-}-E_{+}$ \cite{white-1998a}. (b) A pairing
potential was applied to the left chain only, $\Delta _{L}>0$, while $\Delta
_{R}=0$. The induced pair field
\begin{equation}
\phi _{R,i}\equiv \left\langle c_{R,i,\uparrow }^{\dagger }c_{R,i,\downarrow
}^{\dagger }\right\rangle
\end{equation}
on the right chain was measured. Its sign indicates the sign of $J$. This is
the method we used in most calculations. Method (a) was applied to a small
number of points in parameter space, and found to produce identical results
to those of method (b) for the sign of $J$.

\begin{figure}[t]
\centering
\includegraphics[width=1.0\textwidth]{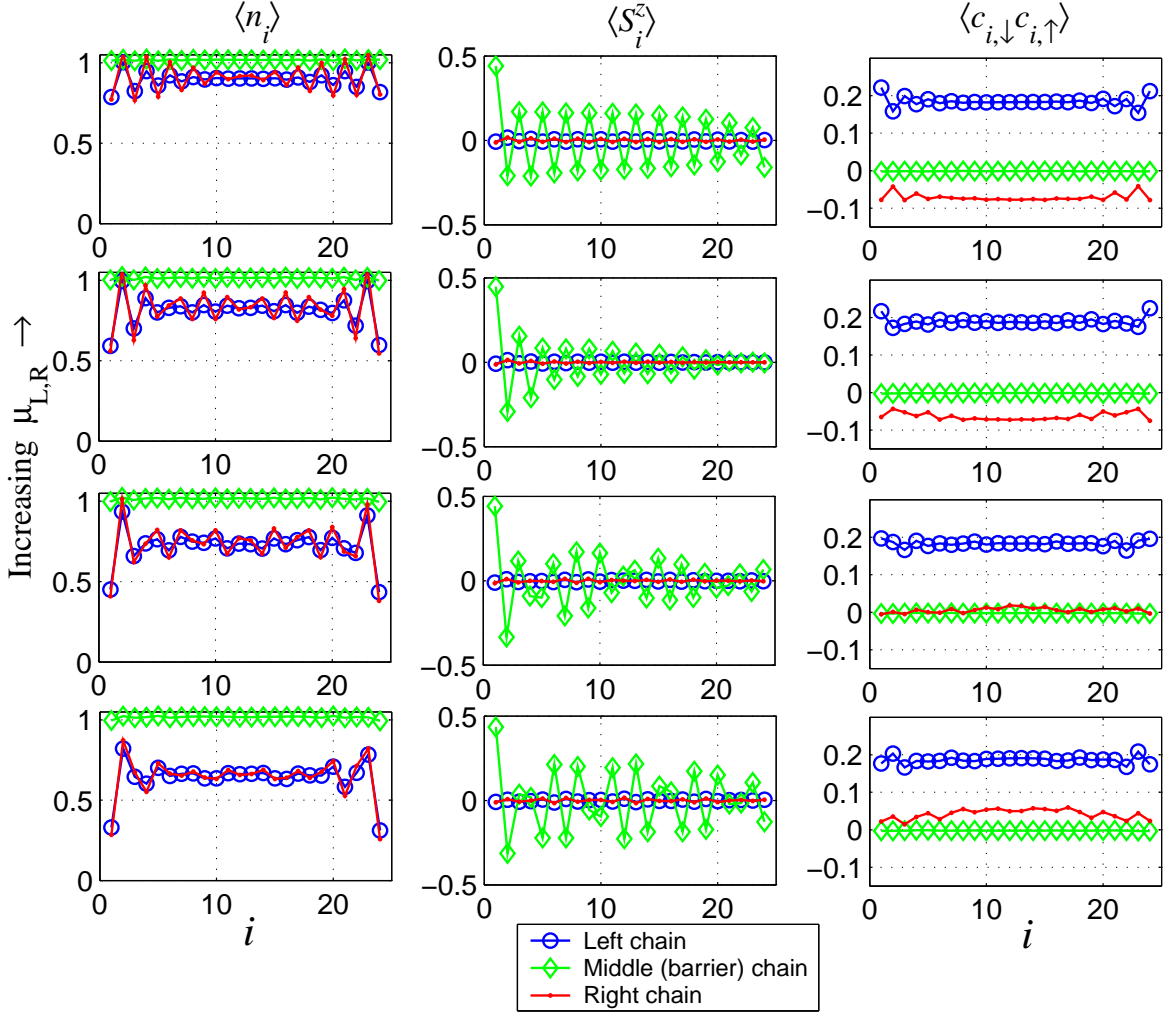}
\caption{(Color online.) The left, middle and right columns show the average
density $\langle n_i\rangle$, $z$ component of the spin $\langle
S^z_i\rangle $ and pair field $\langle c_{i,\downarrow}c_{i,\uparrow}\rangle$, 
respectively, as a function of position $i$ along the chains, calculated
by DMRG for $3\times24$ systems. Circles, diamonds and dots refer to the
left, middle and right chains, respectively. The attractive interaction on
the superconducting (left and right) chains is $|U_{L,R}\vert=2.5$ in all
calculations. A pairing term [Eq. (\protect\ref{Hpair})] was applied with $\Delta_L=0.1$ and $\Delta_R=0$. 
The other model parameters are given by: $t=t^{\prime }=1$, $\protect\mu _{B}=6$, $U_{B}=10$. Each row corresponds to
a single calculation with a specific value of the chemical potential $\protect\mu_{L,R}$ (and hence a particular particle density) on the
superconducting chains.}
\label{fig:DMRG}
\end{figure}

Fig. \ref{fig:DMRG} shows the local expectation values of the particle
number, spin and pair field operators along the three chains for $|U_{L,R}|=2.5$ and various values of $\mu _{LR}$. 
The density of electrons
on the left and right chains increases as $\mu _{LR}$ increases, while the
density on the middle chain is kept close to one particle per site. A
positive pair potential of strength $\Delta _{L}=0.1$ was applied on the
left chain, inducing a positive pair field $\phi _{L}=\langle
c_{L,i,\uparrow }^{\dagger }c_{L,i,\downarrow }^{\dagger }\rangle>0$, while $\Delta _{R}=0$. 
A negative induced pair field $\phi _{R}$ on the right chain
indicates that the effective Josephson coupling $J$ between the left and
right chains is negative. Note that $J$ is negative for the two upper rows
(in which $\langle n_{L,R}\rangle =0.9$, $0.83$ respectively), while for the
two lower rows (where $\langle n_{L,R}\rangle =0.75$, $0.66$) it becomes
positive. This is in agreement with our expectation, based on the
perturbative analysis of the previous subsection, that when the
superconducting chains are close to half filling ($\langle n\rangle =1$),
the negative $J_{2} $ term dominates and the overall Josephson coupling is
more likely to become negative.

The middle column in Fig. \ref{fig:DMRG} shows the expectation value of the $z$ 
component of the spin along the three chains. In order to visualize the
spin correlations, a Zeeman field of strength $h=0.5$ was applied to the $i=1 $ site of the middle chain. 
The results clearly indicate that the two
outer chains have a spin gap (and therefore have a very small induced
moment), while in the half filled middle chain there are strong
antiferromagnetic correlations. Interestingly, as the chemical potential on
the outer chains is decreased, the spin correlations along the middle chain
become incommensurate. This seems to occur at the same point where the
Josephson coupling changes sign (between the second and third row in Fig. \ref{fig:DMRG}). 
This phenomenon was observed for other values of $U_{L,R}$,
as well. The incommensurate correlations can be explained by the
further-neighbor Ruderman-Kittel-Kasuya-Yosida (RKKY)-like interaction which
are induced in the middle chain by the proximity of the outer chains. Upon
decreasing the inter-chain hopping $t^{\prime }$ to 0.7, the spin
correlations in the middle chain become commensurate over the entire range
of $\mu_{L,R}$ (and the region of negative $J$ increases). Why $J>0$ seems
to be favored by incommensurate correlations in the middle chain is not
clear at present.

\begin{figure}[t]
\centering
\includegraphics[width=12cm]{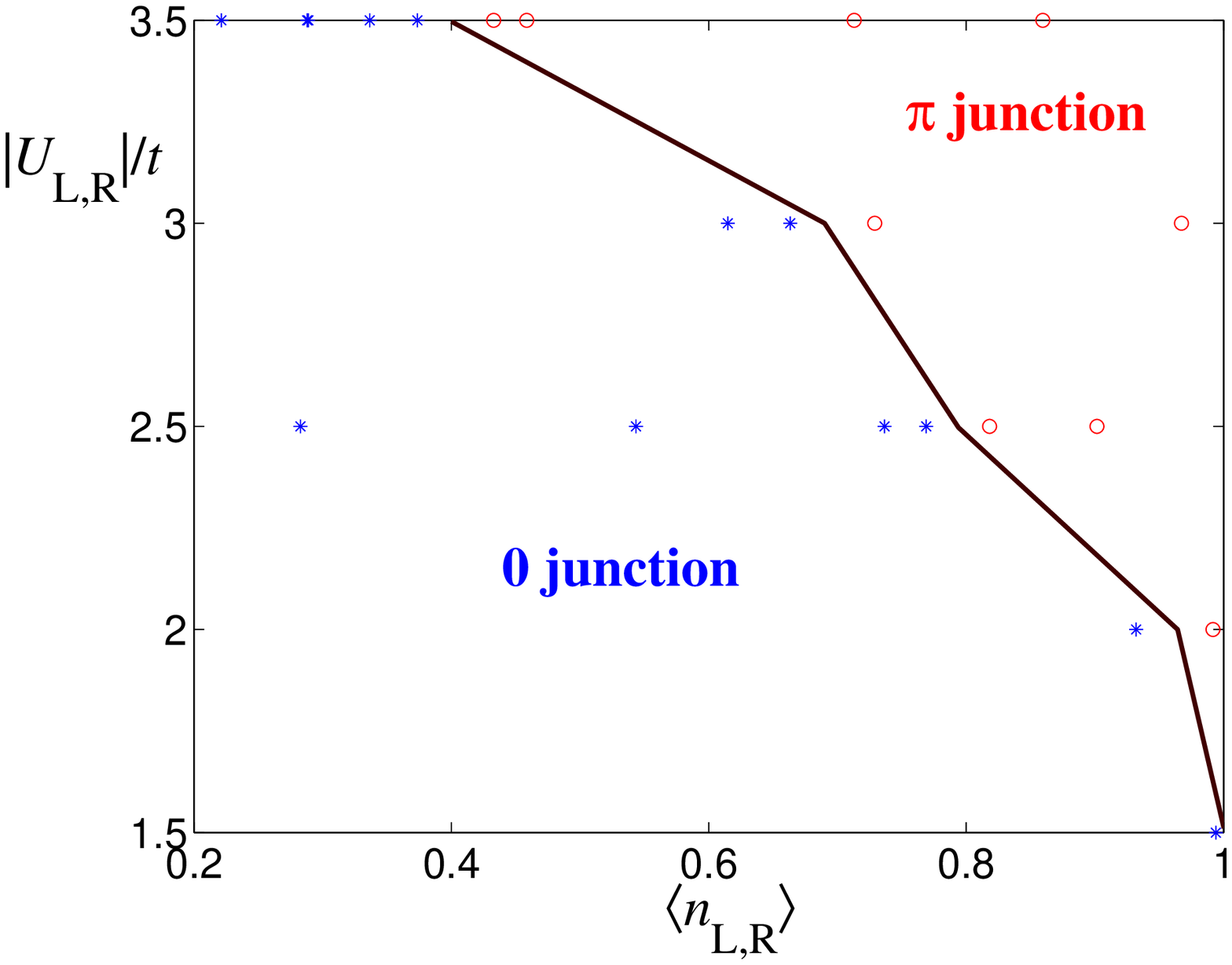}
\caption{(Color online.) Phase diagram of the three chain model [Eq. (\protect\ref{H0},\protect\ref{Hp})] 
from DMRG, as a function of $\vert
U_{L,R}\vert$, the attraction on the left and right chains, and $\langle
n_{L,R}\rangle$, the number of electrons per site on the left and right
chains. The following parameters were used: $t=t^{\prime }=1$, $\protect\mu_{B}=6$, $U_{B}=10$. 
On the middle chain, $\langle n_B \rangle \approx 1$ in
all cases. The symbols show the points that were simulated.}
\label{fig:phase_diagram}
\end{figure}

Fig. \ref{fig:phase_diagram} shows the phase diagram of the three chain
model as a function of the density $\left\langle n_{L,R}\right\rangle $ and
the attractive interaction $\left\vert U_{L,R}\right\vert $ on the outer
chains. In agreement with the perturbative considerations, proximity to $\left\langle n_{L,R}\right\rangle =1$ and large $\left\vert
U_{L,R}\right\vert $ (compared to the bandwidth $4t$) both favor a negative
Josephson coupling between the outer chains.

\subsection{Extension to an infinite array of coupled chains}

\label{sec:model-striped_sc}

The model presented in the previous subsection includes only a single
extended $\pi$ junction. However, it is straightforward to extend this model
to an infinite number of coupled chains with alternating $U$. So long as the
Josephson coupling across a single junction is small, we expect that the
extension to an infinite number of chains will not change it by much.
Therefore, in the appropriate parameter regime in Fig. \ref{fig:phase_diagram}, 
the superconducting order parameter changes sign from
one superconducting chain to the next, forming a striped superconductor (or
unidirectional PDW).

\begin{figure}[t]
\centering
\includegraphics[width=10cm]{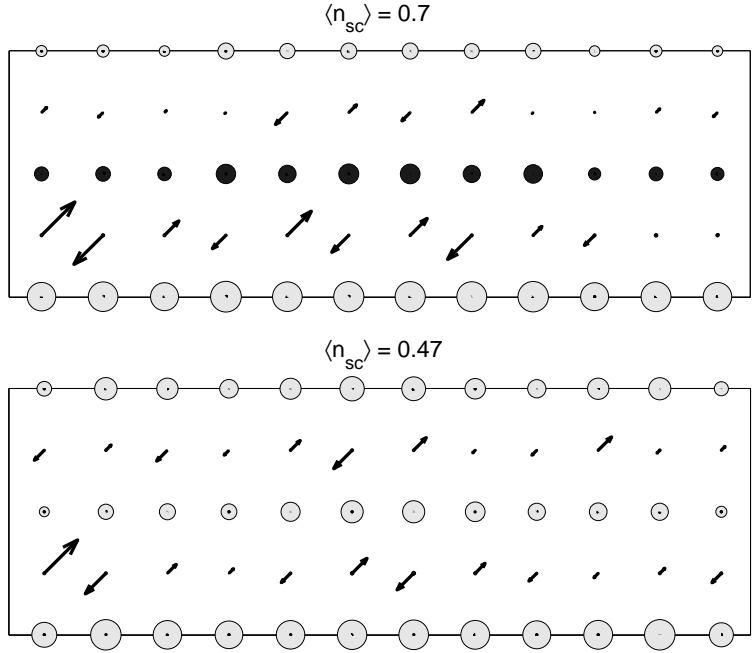}
\caption{The average pair field $\protect\phi=\langle{c^\dagger_\uparrow
c^\dagger_\downarrow}\rangle$ and spin $\langle S^z\rangle$ measured in DMRG
calculations for the $5\times12$ systems with alternating $U$, as described
in the text. The size of the circles indicate the magnitude of $\protect\phi$, 
and their color indicate its sign (bright-positive sign, dark-negative
sign.) The arrows indicate the magnitude and sign of $\langle S^z\rangle$.
In each calculation, a positive pair field $\Delta=0.1$ was applied to the
lower chain, and a Zeeman field $h=0.1$ was applied to the leftmost site of
the second row from the bottom.}
\label{fig:12x5}
\end{figure}

In order to demonstrate that there are no surprises in going from three
chains to two dimensions, we have performed a simulation for a $5\times12$
system composed of $5$ coupled chains with alternating $U=-3,8,-3,8,-3$. As
before, the density of particles on the $U=8$ chains was kept close to $\langle n\rangle=1$, 
making them insulating, while the density of particles
on the $U=-3$ (superconducting) chains was varied. As before, the hopping
parameters are $t=t^{\prime}=1$. A pair field $\Delta=0.1$ was applied on
the bottom superconducting chain, and the induced superconducting order
parameter was measured across the system. Up to $m=2300$ states were kept.
Fig. \ref{fig:12x5} shows the induced pair fields and the expectation value
of $S^z$ throughout the system in two simulations, in which the average
density of particles on the superconducting chains was $\langle
n_{sc}\rangle=0.7$, $0.47$. As expected according to the phase diagram in
Fig. \ref{fig:phase_diagram}, in the $\langle n_{sc}\rangle=0.7$ case the
order parameter changes sign from one superconducting chain to the next,
while in the $\langle n_{sc}\rangle=0.47$ case the sign is uniform. It
therefore seems very likely that under the right conditions, the two
dimensional alternating chain model forms a striped superconductor.

\subsection{Quasiparticle spectrum of a striped superconductor}

\label{sec:qp-spectrum}

The quasiparticle spectrum of a uniform superconductor is typically either
fully gapped, or gapless only on isolated nodal points (or nodal lines in
3D). This is a consequence of the fact that, due to time reversal symmetry,
the points $\mathbf{k}$ and $-\mathbf{k}$ have the same energy. Since the
order parameter carries zero momentum, any point on the Fermi surface is
thus perfectly nested with its time reversed counterpart, and is gapped
unless the gap function $\Delta_\textbf{k}$ vanishes at that point.

For a striped superconductor, the situation is different. Since the order
parameter has non-zero momentum $\mathbf{Q}$, only points that satisfy the
nesting condition $\varepsilon_\textbf{k}=\varepsilon_{-\mathbf{k}+\mathbf{Q}}$, 
where $\varepsilon_\textbf{k}$ is the single particle energy, are gapped
for an infinitesimally weak order. Therefore, generically there are portions
of the Fermi surface that remain gapless \cite{berg-2008}. This is similar
to the case of a CDW or SDW, which generically leave parts of the
(reconstructed) Fermi surface gapless, until the magnitude of the order
parameter reaches a certain critical value. The spectral properties of a
striped superconductor where studied in detail in Refs. \cite{dror,yang-2008b}.

\begin{figure}[t]
\centering
\includegraphics[width=1.0\textwidth]{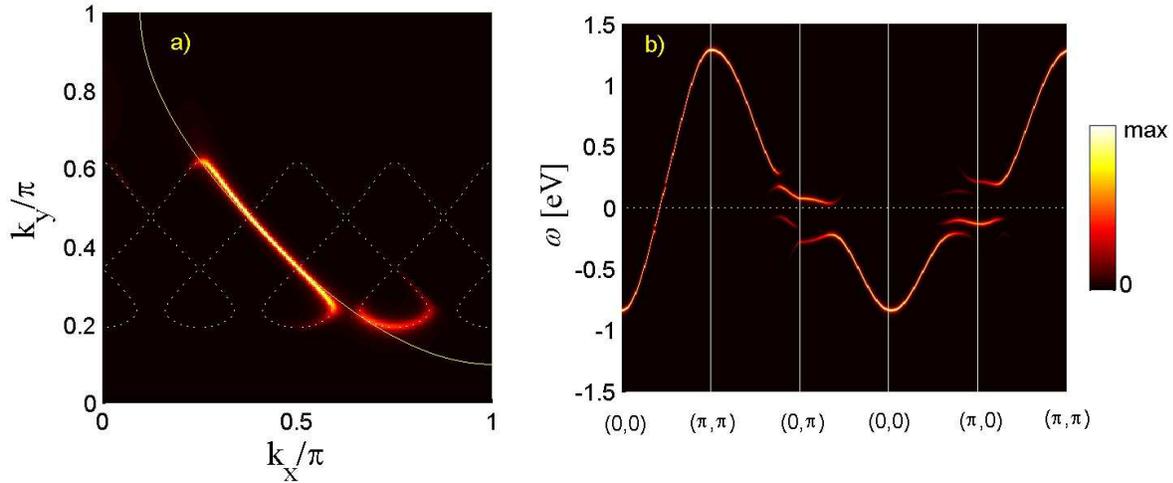}
\caption{(Color online.) (a) The spectral function $A(\mathbf{k},\protect\omega=0)$ for a striped superconductor. 
The band parameters used in the
calculation where fitted to the ARPES spectrum of LSCO \protect\cite{shen08}: $t=0.25$, $t^{\prime}=-0.031863$, $t^{\prime\prime}=0.016487$, $t^{\prime\prime\prime}=0.0076112$, where $t$, $t^{\prime}$,... are nearest
neighbor hopping, second-nearest neighbor hopping and so on, chemical
potential $\protect\mu=-0.16235$. (All the parameters above are measured in eV.) The striped superconducting order parameter has a wavevector of $\mathbf{Q}=(2\protect\pi/8, 0)$, and its magnitude is $\Delta_\textbf{Q}=60$meV. The order parameter is of ``d-wave character'', in the sense that it is
of opposite sign on $x$ and $y$ oriented bonds. The thin solid line shows
the underlying bare Fermi surface, and the dotted line shows the Fermi
surface in the presence of the PDW. (b) $A(\mathbf{k},\protect\omega)$ for
the same model parameters along a cut in k-space.}
\label{fig:spectrum}
\end{figure}

As an illustration, we present in Fig. \ref{fig:spectrum} the spectral
function $A(\mathbf{k},\omega =0)$ of a superconductor with band parameters
fitted to the ARPES spectrum of LSCO \cite{shen08} and a striped
superconducting order parameter with a single wavevector $\mathbf{Q}=(2\pi
/8,0)$ of magnitude $\Delta _{\mathbf{Q}}=60$meV. Note that a portion of the
Fermi surface around the nodal (diagonal) direction remains ungapped (a
\textquotedblleft Fermi arc\textquotedblright\ \cite{Norman, Kanigel}),
while both antinodal directions [around $(\pi ,0)$ and $(0,\pi )$] are
gapped. The Fermi arc is in fact the back side of a reconstructed Fermi
pocket, but only the back side has a sizable spectral weight \cite{sudip}.
Its length depends on the magnitude of the order parameter: the larger the
magnitude of $\Delta _{\mathbf{Q}}$, the smaller is the arc. Note that $A(\mathbf{k},\omega =0)$ is not symmetric under rotation by $\pi /2$, because
the striped superconducting order breaks rotational symmetry. However, in a
system with an LTT symmetry (such as LBCO near $x=1/8$ doping) both
orientations of stripes are present, and an ARPES experiment would see the
average of the picture in Fig. \ref{fig:spectrum} and its rotation by $\pi/2 $.

\section{Order parameter theory of the PDW state}
\setcounter{footnote}{0}

\label{sec:op}

In this section, we explore the aspects of the theory of a PDW that can be
analyzed without reference to microscopic mechanisms. We focus on the
properties of ordered states at $T=0$, far from the point of any quantum
phase transition, where for the most part fluctuation effects can be
neglected. (The one exception to the general rule is that, where we discuss
effects of disorder, we will encounter various spin-glass related phases
where fluctuation effects, even at $T=0$, can qualitatively alter the
phases.) For simplicity, most of our discussion is couched in terms of a
Landau theory, in which the effective free energy is expanded in powers of
the order parameters; this is formally \textit{not} justified deep in an
ordered phase, but it is a convenient way to exhibit the consequences of the
order parameter symmetries.

\subsection{Order parameters and symmetries}

\label{sec:symmetries}

We will now define the various order parameters introduced in this section
and discuss their symmetry properties. The striped superconducting order
parameter $\Delta _{\mathbf{Q}}$ is a charge $2e$ complex scalar field,
carrying momentum $\mathbf{Q}$. To define it microscopically, we write the
superconducting order parameter as

\begin{eqnarray}
\phi \left( \mathbf{r},\mathbf{r}^\prime \right) &\equiv &\left\langle \psi
_{\uparrow }^{\dagger }\left(\mathbf{r}\right) \psi _{\downarrow }^{\dagger
}\left( \mathbf{r}^\prime \right) \right\rangle  \nonumber \\
&=&F\left( \mathbf{r}-\mathbf{r}^\prime \right) \left[ \Delta _{0}+\Delta _{\mathbf{Q}}e^{i\mathbf{Q}\cdot \mathbf{R}}+ \Delta_{-\mathbf{Q}}
e^{-i\mathbf{Q}\cdot \mathbf{R}}\right],  \nonumber \\
&&
\end{eqnarray}
where $\mathbf{R}=\left( \mathbf{r}+\mathbf{r}^\prime \right) /2$, $F\left(\mathbf{r}-\mathbf{r}^\prime \right) $ 
is some short range function (for a
``d-wave-like'' striped superconductor, $F\left( \mathbf{r}\right) $ changes
sign under $90^\circ$ rotation), and $\Delta _{0}$ is the uniform 
${\bf Q} = {\bf 0}$    component
of the order parameter.\footnote{A state in which both components of the SC order parameter {\em coexist}, $\Delta_0\neq 0$ {\em and} $\Delta_{\textbf{Q}}\neq 0$  is certainly not ``uniform''. Even a weak $\Delta_{\textbf{Q}}\neq 0$ implies the existence of a modulation  of the local amplitude of the SC order parameter, and a SC state is ``truly uniform'' only if $\Delta_{\textbf{Q}}=0$. Nevertheless, as we will see in Section \ref{sec:uniform}, the properties of a SC state in which both order parameters coexist are largely dominated by the ``uniform'' component $\Delta_0$, and the striking features of the PDW state are not 
directly
observable. In this sense, the uniform-PDW coexisting SC state is effectively ``uniform.''} 
In the rest of this subsection, we set $\Delta_{0}=0$. The effect of $\Delta _{0} $ is discussed in subsection \ref{sec:uniform}.
To be concrete, we assume that the host crystal is tetragonal, and that
there are therefore two potential symmetry related ordering wave vectors, $\mathbf{Q}$ and $\bar{\mathbf{Q}}$, which are mutually orthogonal, so $\Delta_{\bar{\mathbf{Q}}}$ must be treated on an equal footing with $\Delta
_{\mathbf{Q}}$. (The discussion is easily generalized to crystals with other
point-group symmetries.) Similarly, for simplicity, spin-orbit coupling is
assumed to be negligible.

The order parameters that may couple to $\Delta _{\mathbf{Q}}$ and their
symmetry properties are as follows: The nematic order parameter $N$ is a
real pseudo-scalar field; the CDW $\rho _{\mathbf{K}}$ with $\mathbf{K}=2\mathbf{Q}$ is a scalar field; $\vec{S}_{\mathbf{Q}}$ is a neutral
spin-vector field. All these order parameters are electrically neutral.
Under spatial rotation by $\pi /2$, $N\rightarrow -N$, $\rho _{\mathbf{K}}\rightarrow \rho _{\bar{\mathbf{K}}}$, $\vec{S}_{\mathbf{Q}}\rightarrow
\vec{S}_{\bar{\mathbf{Q}}}$, and $\Delta _{\mathbf{Q}}\rightarrow \pm \Delta_{\bar{\mathbf{Q}}}$, where $\pm$ refers to a d-wave or s-wave version of
the striped superconductor. Under spatial translation by $\mathbf{r}$, $N\rightarrow N$, $\rho_{\mathbf{K}}\rightarrow e^{i\mathbf{K}\cdot \mathbf{r}}\rho _{\mathbf{K}}$, $\vec{S}_{\mathbf{Q}}\rightarrow e^{i\mathbf{Q}\cdot
\mathbf{r}}\vec{S}_{\mathbf{Q}}$, and $\Delta_{\mathbf{Q}}\rightarrow e^{i\mathbf{Q}\cdot \mathbf{r}}\Delta _{\mathbf{Q}}$. Note that since the SDW
and CDW orders are real, $\vec{S}^\star_\mathbf{Q}=\vec{S}_\mathbf{-Q}$ and $\rho^\star_\mathbf{K}=\rho_\mathbf{-K}$. Generally, $\Delta_\mathbf{Q}$ and $\Delta^\star_\mathbf{Q}$ are independent.

\subsection{Landau Theory}

\label{sec:landau}

Specifically, the emphasis in this section is on the interrelation between
striped superconducting order and other orders. There is a necessary
relation between this order and CDW and nematic (or orthorhombic) order,
since the striped superconductor breaks both translational and rotational
symmetries of the crystal. From the microscopic considerations, above, and
from the phenomenology of the cuprates, we also are interested in the
relation of superconducting and SDW order.
The Landau effective free energy density can then be expanded in powers of
these fields:
\begin{equation}
\mathcal{F}=\mathcal{F}_{2}+\mathcal{F}_{3}+\mathcal{F}_{4}+\ldots
\label{F}
\end{equation}
where $\mathcal{F}_{2}$, the quadratic term, is simply a sum of decoupled
terms for each order parameter,
\begin{eqnarray}
\mathcal{F}_{3} &=&\gamma _{s}[\rho _{-\mathbf{K}}\vec{S}_{\mathbf{Q}}\cdot
\vec{S}_{\mathbf{Q}} + \rho_{-{\bar{\mathbf{K}}}}\vec{S}_{\bar{\mathbf{Q}}}\cdot \vec{S}_{\bar{\mathbf{Q}}}+\mathrm{c.c.}] \\
&&+\gamma _{\Delta }[\rho _{-\mathbf{K}}\Delta _{-\mathbf{Q}}^{\star
}\Delta_{\mathbf{Q}} +\rho_{-\bar{\mathbf{K}}}\Delta _{-\bar{\mathbf{Q}}}^{\star }\Delta_{\bar{\mathbf{Q}}}+\mathrm{c.c.}]  \nonumber \\
&&+g_{\Delta }N[\Delta _{\mathbf{Q}}^{\star }\Delta _{\mathbf{Q}}+ \Delta_{-\mathbf{Q}}^{\star }\Delta _{-\mathbf{Q}}-\Delta _{\bar{\mathbf{Q}}}^{\star
} \Delta_{\bar{\mathbf{Q}}}-\Delta _{-\bar{\mathbf{Q}}}^{\star}\Delta _{-\bar{\mathbf{Q}}}]  \nonumber \\
&&+g_{s}N[\vec{S}_{-\mathbf{Q}}\cdot \vec{S}_{\mathbf{Q}}-\vec{S}_{-\bar{\mathbf{Q}}}\cdot \vec{S}_{\bar{\mathbf{Q}}}]  \nonumber \\
&&+g_{c}N[\rho _{-\mathbf{K}}\rho _{\mathbf{K}}-\rho _{-\bar{\mathbf{K}}}\rho _{\bar{\mathbf{K}}}],  \nonumber
\end{eqnarray}
and the fourth order term, which is more or less standard, is shown
explicitly below. 

The effect of the cubic term proportional to $\gamma _{s}$ on the interplay
between the spin and charge components of stripe order has been analyzed in
depth in \cite{zke}. Similar analysis can be applied to the other terms. In
particular, the $\gamma_\Delta$ and $g_\Delta$ terms imply\footnote{Note that the $\gamma_\Delta$ term is odd under a particle-hole
transformation, which takes $\rho_\mathbf{K}\rightarrow -\rho_\mathbf{K}$.
Therefore, if the system has exact particle-hole symmetry, this term
vanishes, and there is no necessary relation between $\Delta_\mathbf{Q}$ and
$\rho_\mathbf{K}$. Microscopic systems are generically not symmetric under
charge conjugation. However, some real systems (\emph{e.g.} the cuprates)
are not too far from being particle-hole symmetric, and therefore in these
systems $\gamma_\Delta$ is expected to be relatively small.} that the
existence of superconducting stripe order ($\Delta _{\mathbf{Q}}\neq 0$, and $\Delta _{\bar{\mathbf{Q}}}=0$), implies the existence of nematic order ($N\neq 0$) and charge stripe order with half the period ($\rho _{2\mathbf{Q}}\neq 0$). However, the converse statement is not true: while CDW order with
ordering wave vector $2\mathbf{Q}$ or nematic order tend to promote PDW
order, depending on the magnitude of the quadratic term in $\mathcal{F}_{2}$, PDW order may or may not occur.

One new feature of the coupling between the PDW and CDW order is that it
produces a sensitivity to disorder which is not normally a feature of the
superconducting state. In the presence of quenched disorder, there is always
some amount of spatial variation of the charge density, $\rho(\mathbf{r})$,
of which the important portion for our purposes can be thought of as being a
pinned CDW, that is, a CDW with a phase which is a pinned, slowly varying
function of position, $\rho(\mathbf{r})=|\rho _{\mathbf{K}}|\cos [\mathbf{K}\cdot \mathbf{r}+\phi (\mathbf{r})]$. Below the nominal striped
superconducting ordering temperature, we can similarly express the PDW order
in terms of a slowly varying superconducting phase, $\Delta(\mathbf{r})=
|\Delta _{\mathbf{Q}}|\exp [i\mathbf{Q}\cdot r+i\theta _{\mathbf{Q}}(\mathbf{r})]+
|\Delta_{-\mathbf{Q}}|\exp[-i\mathbf{Q}\cdot \mathbf{r}+i\theta_{-\mathbf{Q}}(\mathbf{r})]$. The resulting contribution to $\mathcal{F}_{3}$
is
\begin{equation}
\mathcal{F}_{3,\gamma }=2\gamma _{\Delta }|\rho _{\mathbf{K}}\Delta _{\mathbf{Q}}\Delta _{-\mathbf{Q}}|\cos [2\theta _{-}(\mathbf{r})-\phi(\mathbf{r})].  
\label{F3gamma}
\end{equation}
where
\begin{eqnarray}
\theta _{\pm }(\mathbf{r}) &\equiv &[\theta _{\mathbf{Q}}(\mathbf{r})\pm
\theta _{-\mathbf{Q}}(\mathbf{r})]/2;  \label{thetapm} \\
\theta _{\pm \mathbf{Q}}(\mathbf{r}) &=&[\theta _{+}(\mathbf{r})\pm \theta
_{-}(\mathbf{r})] .  \nonumber
\end{eqnarray}
The aspect of this equation that is notable is that the disorder couples
directly to a piece of the superconducting phase, $\theta_{-}$. No such
coupling occurs in usual $0$ momentum superconductors.

It is important to note that the condition that $\Delta(\mathbf{r})$ be
single valued implies that $\theta_{ \mathbf{Q}}(\mathbf{r})$ and $\theta_{ -\mathbf{Q}}(\mathbf{r})$ are defined modulo $2\pi$. Correspondingly, $\theta_\pm$ are defined modulo $\pi$, subject to the constraint that if $\theta_{\pm} \to \theta_{\pm} + \pi m_\pm$ then $m_++m_- $ must be an even
integer. Since $\phi$ and $\theta_-$ are locked to each other at long
distances, the possible topological excitations of the coupled PDW-CDW
system are thus point defects in 2D and line defects in 3D classified by the
circulation of $\theta_+$ and $\phi$ on any enclosing contour. The
elementary topological defects thus are: a) An ordinary superconducting
vortex, about which $\Delta \theta_+=2\pi$ and $\Delta \phi=0$. b) A
bound-state of a half vortex and a dislocation,\footnote{The possibility of half vortices in a striped superconductor and their effect on the phase diagram in the clean case was discussed by D. F. Agterberg and H. Tsunetsugu\cite{agterberg08}.} 
about which $\Delta
\theta_+=\pi$ and $\Delta \phi=2\pi$. c) A double dislocation (or
dislocation bound state) about which $\Delta \theta_+=0$ and $\Delta
\phi=4\pi$. All these defects have a logarithmically divergent energy in 2D,
or energy per unit length in 3D; the prefactor of the logarithm is
determined by the superfluid stiffness for the vortex, the elastic modulus
of the CDW for the double vortex, and an appropriate sum of these two
stiffnesses for the half vortex. Consequences of this rich variety of
topological defects are discussed in \cite{berg-2007,Radzihovsky-2008,berg-2009}

An important consequence of the coupling between the superconducting and CDW
phase is that the effect of quenched disorder, as in the case of the CDW
itself, destroys long-range superconducting stripe order. (This statement is
true \cite{larkin}, even for weak disorder, in dimensions $d<4$.) Naturally,
the way in which this plays out depends on the way in which the CDW state is
disordered.

In the most straightforward case, the CDW order is punctuated by random,
pinned dislocations, \textit{i.e.} $2\pi$ vortices of the $\phi$ field. The
existence of the coupling in Eq.~\ref{F3gamma} implies that there must be an
accompanying $\pi$ vortex in $\theta_{-}$. The condition of
single-valued-ness implies that there must also be an associated half-vortex
or anti-vortex in $\theta_{+}$. 
If these latter vortices are fluctuating, they destroy the superconducting
state entirely, leading to a resistive state with short-ranged striped
superconducting correlations. If they are frozen, the resulting state is
analogous to the ordered phase of an $XY$ spin-glass: such a state has a
non-vanishing Edwards-Anderson order parameter, spontaneously breaks
time-reversal symmetry, and, presumably, has vanishing resistance but no
Meissner effect and a vanishing critical current. In 2D, according to
conventional wisdom, a spin-glass phase can only occur at $T=0$, but in 3D
there can be a finite temperature glass transition \cite{youngreview}.

In 3D there is also the exotic possibility that, for weak enough quenched
disorder, the CDW forms a Bragg-glass phase, in which long-range order is
destroyed, but no free dislocations occur \cite{giamarchi,daniel,gingras}.
In this case, $\phi$ can be treated as a random, but single-valued function
- correspondingly, so is $\theta_{-}$. The result is a superconducting
Bragg-glass phase which preserves time reversal symmetry and, presumably,
acts more or less the same as a usual superconducting phase. It is believed
that a Bragg-glass phase is not possible in 2D \cite{daniel}.

Another perspective on the nature of the superconducting state can be
obtained by considering a composite order parameter which is proportional to
$\Delta_Q\Delta_{-Q}$. There is a cubic term which couples a uniform, charge
$4e$ superconducting order parameter, $\Delta_4$, to the PDW order:
\begin{equation}
\mathcal{F}_3^\prime =g_4 \{\Delta_4^\star [\Delta_\vecQ \Delta_{-\mathbf{Q}} + \Delta_{\bar \mathbf{Q}} \Delta_{-\bar \mathbf{Q}}] +\mathrm{c.c.}\}
\end{equation}
This term implies that whenever there is PDW order, there is also
necessarily charge $4e$ uniform superconducting order. However, since this
term is independent of $\theta_-$, it would be totally unaffected by
Bragg-glass formation of the CDW. The half-vortices in $\theta_+$ discussed
above can simply be viewed as the fundamental ($hc/4e$) vortices of a charge
$4e$ superconductor.

Some additional physical insight can be gained by examining the
quartic terms ($\mathcal{F}_{4}$ in Eq. \ref{F}). Let us write all
the possible fourth order terms
consistent with symmetry: 
\begin{eqnarray}
\mathcal{F}_{4} &=&u\left( \vec{S}_{\mathbf{Q}}\cdot \vec{S}_{\mathbf{Q}}\Delta _{\mathbf{Q}}^{\star }\Delta _{-\mathbf{Q}}+
\vec{S}_{\bar{\mathbf{Q}}}\cdot \vec{S}_{\bar{\mathbf{Q}}}\Delta _{\bar{\mathbf{Q}}}^{\star}\Delta_{-\bar{\mathbf{Q}}}+\mathrm{c.c.}\right)  
\nonumber \\
&&+\left( v_{+}[\vec{S}_{-\mathbf{Q}}\cdot \vec{S}_{\mathbf{Q}}+
\vec{S}_{-\bar{\mathbf{Q}}}\cdot \vec{S}_{\bar{\mathbf{Q}}}]+\tilde{v}_{+}[|\rho _{\mathbf{K}}|^{2}+|\rho _{\bar{\mathbf{K}}}|^{2}]\right)  \nonumber \\
&&\ \ \ \ \ \ \times \left( |\Delta _{\mathbf{Q}}|^{2}+|\Delta _{-\mathbf{Q}}|^{2}+
|\Delta _{{\bar{\mathbf{Q}}}}|^{2}+|\Delta _{-{\bar{\mathbf{Q}}}}|^{2}\right\}  
\nonumber \\
&&+\left( v_{-}[\vec{S}_{-\mathbf{Q}}\cdot \vec{S}_{\mathbf{Q}}-
\vec{S}_{-\bar{\mathbf{Q}}}\cdot \vec{S}_{\bar{\mathbf{Q}}}]+\tilde{v}_{-}[|\rho_{\mathbf{K}}|^{2}-|\rho _{\bar{\mathbf{K}}}|^{2}]\right)  
\nonumber \\
&&\ \ \ \ \ \ \times \left( |\Delta _{\mathbf{Q}}|^{2}+
|\Delta _{-\mathbf{Q}}|^{2}-|\Delta _{{\bar{\mathbf{Q}}}}|^{2}-|\Delta _{-{\bar{\mathbf{Q}}}}|^{2}\right)  
\nonumber \\
&&+vN^{2}\left\{ \left( |\Delta _{\mathbf{Q}}|^{2}+
|\Delta _{-\mathbf{Q}}|^{2}\right) +\left( |\Delta _{\bar{\mathbf{Q}}}|^{2}+|\Delta _{-\bar{\mathbf{Q}}}|^{2}\right) \right\}  
\nonumber \\
&&+\lambda _{+}\left\{ \left( |\Delta _{\mathbf{Q}}|^{2}+|\Delta _{-\mathbf{Q}}|^{2}\right) ^{2}+
\left( |\Delta _{\bar{\mathbf{Q}}}|^{2}+|\Delta _{-\bar{\mathbf{Q}}}|^{2}\right) ^{2}\right\}  
\nonumber \\
&&+\lambda _{-}\left\{ \left( |\Delta _{\mathbf{Q}}|^{2}-|\Delta _{-\mathbf{Q}}|^{2}\right) ^{2}+
\left( |\Delta _{\bar{\mathbf{Q}}}|^{2}-|\Delta _{-\bar{\mathbf{Q}}}|^{2}\right) ^{2}\right\}  
\nonumber \\
&&+\lambda
(|\Delta_{\mathbf{Q}}|^2+|\Delta_{-\mathbf{Q}}|^2)(|\Delta_{\bar{\mathbf{Q}}}|^2+|\Delta_{-\bar{\mathbf{Q}}}|^2)
  \nonumber \\
&&+\ldots \label{F4}
\end{eqnarray}
where we have explicitly shown all the terms involving $\Delta _{\mathbf{Q}}$, while the terms $\ldots $ represent the remaining quartic terms
all of which, with the exception of those involving $N$, are
exhibited explicitly in \cite{zke}.

There are a number of features of the ordered phases which depend
qualitatively on the sign of various couplings. Again, this is
very similar to what happens in the case of CDW order - see, for
example, \cite{robertson, DelMaestro-2006}. For instance,
depending on the sign of $\lambda$, either unidirectional
(superconducting stripe) or bidirectional (superconducting
checkerboard) order is favored.

On physical grounds, we have some information concerning the sign
of various terms in $\mathcal{F}_{4}$. The term proportional to
$u$ determines the relative phase of the spin and superconducting
stripe order---we believe $ u>0 $ which thus favors a $\pi/2$
phase shift between the SDW and the striped superconducting order,
\textit{i.e.}\ the peak of the superconducting order occurs where
the spin order passes through zero. The other interesting thing
about this term is that it implies an effective cooperativity
between spin and striped superconducting order. The net effect,
\textit{i.e.}\ whether spin and striped superconducting order
cooperate or fight, is determined by the sign of
$|u|-v_{+}-v_{-}$, such
that they cooperate if $|u|-v_{+}-v_{-}>0 $ and oppose each other if $|u|-v_{+}-v_{-}<0$. It is an interesting possibility that spin
order and superconducting stripe order can actually favor each
other even with all ``repulsive'' interactions. The term
proportional to $\lambda _{-}$
determines whether the superconducting stripe order tends to be real ($\lambda _{-}>0$), with a superconducting order that simply changes
sign as a function of position, or a complex spiral, which
supports ground-state currents ($\lambda _{-}<0$).

\subsection{Coexisting uniform and striped order parameters}

\label{sec:uniform}

Finally, we comment on the case of coexisting uniform and striped
superconducting order parameters. Such a state is not thermodynamically
distinct from a regular (uniform) superconductor coexisting with a charge
density wave, even if the uniform superconducting component is in fact
weaker than the striped component. Therefore, we expect many of the special
features of the striped superconductor (such as its sensitivity to potential
disorder) to be lost. Here, we extend the Landau free energy to include a
uniform superconducting component, and show that this is indeed the case.

We will now analyze the coupling of a striped superconducting order
parameter $\Delta _{\mathbf{Q}}$ to a uniform order parameter, $\Delta _{0}$. In this case, we have to consider in addition to the order parameters
introduced in Sec. \ref{sec:op} a CDW order parameter with wavevector $\mathbf{Q}$, denoted by $\rho _{\mathbf{Q}}$. The additional cubic terms in
the Ginzburg-Landau free energy are 
\begin{eqnarray}
\mathcal{F}_{3,u} &=&\gamma _{\mathbf{Q}}\Delta _{0}^{\star }\left[ \rho _{\mathbf{Q}}\Delta _{-\mathbf{Q}}+\rho _{-\mathbf{Q}}\Delta _{\mathbf{Q}}+\rho _{\mathbf{\bar{Q}}}\Delta _{-\bar{\mathbf{Q}}}+\rho _{-\bar{\mathbf{Q}}}\Delta _{\bar{\mathbf{Q}}}\right] +\mathrm{c.c.}  \nonumber \\
&&+g_{\rho }\left[ \rho _{-2\mathbf{Q}}\rho _{\mathbf{Q}}^{2}+\rho _{-2\bar{\mathbf{Q}}}\rho _{\bar{\mathbf{Q}}}^{2}+\mathrm{c.c.}\right] .  \label{Fu3}
\end{eqnarray}
Eq.~\ref{Fu3} shows that if both $\Delta _{0}$ and $\Delta _{\mathbf{Q}}$
are non-zero, there is necessarily a coexisting non-zero $\rho _{\mathbf{Q}}$, through the $\gamma _{\mathbf{Q}}$ term. The additional quartic terms
involving $\Delta _{0}$ are
\begin{eqnarray}
\mathcal{F}_{4,u} &=&u_{\Delta }\left( \Delta _{0}^{\star 2}\Delta _{\mathbf{Q}}\Delta _{-\mathbf{Q}}+
\Delta _{0}^{\star 2}\Delta_{\bar{\mathbf{Q}}}\Delta _{-\bar{\mathbf{Q}}}+\mathrm{c.c.}\right) +
\delta |\Delta_{0}|^{2}[|\Delta _{\mathbf{Q}}|^{2}+|\Delta _{\bar{\mathbf{Q}}}|^{2}]
\nonumber \\
&&+|\Delta _{0}|^{2}\left[ u_{\rho }\left( \left\vert \rho _{\mathbf{Q}}\right\vert ^{2}+\left\vert \rho _{\bar{\mathbf{Q}}}\right\vert ^{2}\right)
+u_{\rho }^{\prime }\left( \left\vert \rho _{2\mathbf{Q}}\right\vert
^{2}+\left\vert \rho _{2\bar{\mathbf{Q}}}\right\vert ^{2}\right) \right]
\nonumber \\
&&+v^{\prime }|\Delta _{0}|^{2}[\vec{S}_{-\mathbf{Q}}\cdot \vec{S}_{\mathbf{Q}}+\vec{S}_{-\bar{\mathbf{Q}}}\cdot \vec{S}_{\bar{\mathbf{Q}}}].  \label{Fu4}
\end{eqnarray}
Let us now consider the effect of quenched disorder. Following the
discussion preceding Eq.~\ref{thetapm}, we write the order parameters in
real space as
\begin{equation}
\Delta \left( \mathbf{r}\right) =\left\vert \Delta _{0}\right\vert
e^{i\theta _{0}}+\left\vert \Delta _{\mathbf{Q}}\right\vert e^{i\left(
\theta _{\mathbf{Q}}+\mathbf{Q}\cdot \mathbf{r}\right) }+\left\vert \Delta
_{-\mathbf{Q}}\right\vert e^{i\left( \theta _{-\mathbf{Q}}-\mathbf{Q}\cdot
\mathbf{r}\right) }
\end{equation}
and
\begin{equation}
\rho \left( r\right) =\left\vert \rho _{\mathbf{Q}}\right\vert \cos \left(
\mathbf{Q}\cdot \mathbf{r}+\phi _{\mathbf{Q}}\right) +\left\vert \rho _{2\mathbf{Q}}\right\vert \cos \left( 2\mathbf{Q}\cdot \mathbf{r}+\phi \right) .
\end{equation}
Let us assume that the disorder nucleates a point defect in the CDW, which
in this case corresponds to a $2\pi $ vortex in the phase $\phi _{\mathbf{Q}} $. By the $g_{\rho }$ term in Eq. \ref{Fu3}, this induces a $4\pi $ vortex
in $\phi $. (Note that in the presence of $\rho _{\mathbf{Q}}$, a $2\pi $
vortex in $\phi $ is not possible.) The $\gamma _{\Delta }$ term in Eq.~\eqref{F3gamma} then dictates a $2\pi $ vortex in the phase $\theta _{-}=\left(
\theta _{\mathbf{Q}}-\theta _{-\mathbf{Q}}\right) /2$. However, unlike
before, this vortex does not couple to the global superconducting phase $\theta _{+}=\left( \theta _{\mathbf{Q}}+\theta _{-\mathbf{Q}}\right) /2$.
Therefore, an arbitrarily small uniform superconducting component is
sufficient to remove the sensitivity of a striped superconductor to
disorder, and the system is expected to behave more or less like a regular
(uniform) superconductor, albeit with a modulated amplitude of the order parameter.

Since the usual (uniform) superconducting order and the PDW break distinct
symmetries, nothing can be said, in general, about the conditions in which
they will coexist. However, microscopic considerations can, in some cases,
yield generic statements, too. For example, in a striped SC, a uniform
component of the order parameter can be generated by dimerizing the stripe
order, such that the positive and negative strips of superconducting order
are made alternately broader and narrower. In any structure (such as the LTT
structure of LBCO), in which there is zero Josephson coupling between
neighboring layers, a coupling is generated, thus lowering the energy of the
system, in proportion to the square of the dimerization. Presumably, so long
as the PDW period is incommensurate with the underling lattice, there is
also a quadratic energy cost to dimerization which is related to an
appropriate generalized elastic constant of the PDW. However, if the PDW has
a long period, this elastic constant will be vanishingly small. Thus, any
long period, incommensurate PDW may generically be expected to be unstable
toward the generation of a small amount of uniform SC order.

\section{Non-collinear order and time reversal symmetry breaking}
\setcounter{footnote}{0}
\label{sec:Tbreaking}

\begin{figure}[t]
\centering
\includegraphics[width=1.0\textwidth]{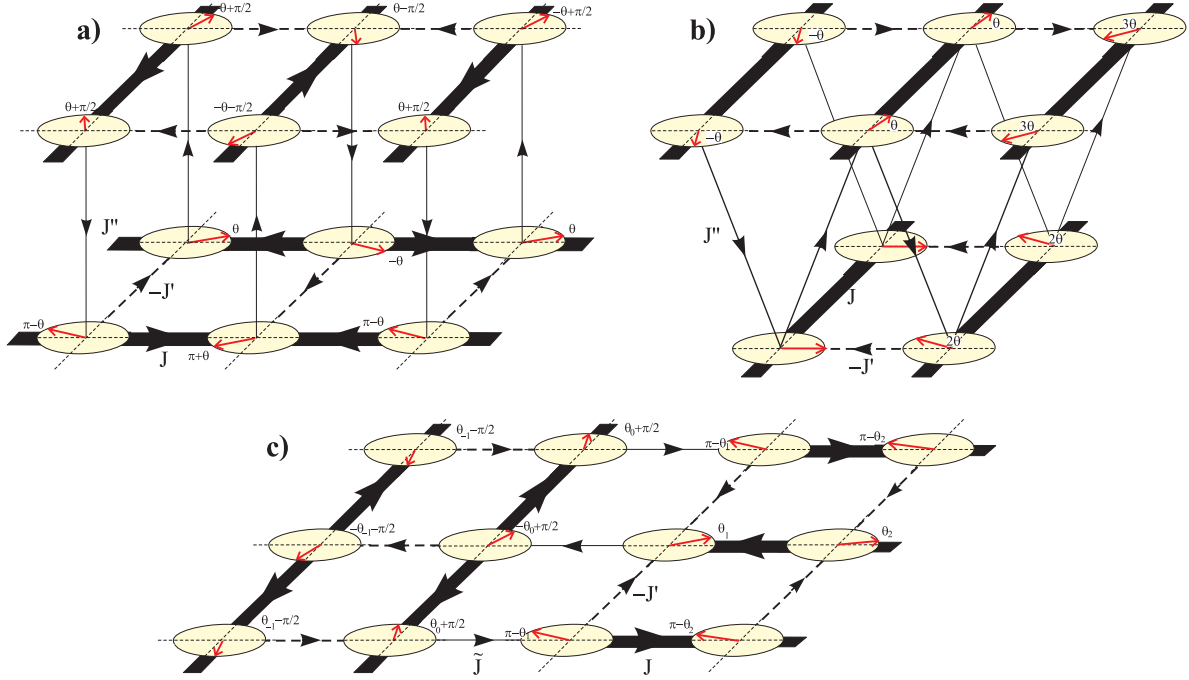}
\caption{(Color online.) (a) Model for a striped superconductor
with an LTT structure. Solid (dashed) lines represent positive
(negative) Josephson couplings. The arrow on the center of each
link indicates the direction of the equilibrium current across
that link. The red arrows on the vertices represent the
superconducting phases. (b) Same as (a) for an orthorhombic
striped superconductor, where the charge stripes are shifted by
half a period from one layer to the next. (c) An in-plane domain
wall.} \label{fig:spiral}
\end{figure}

In a layered system, PDW\ order in the planes can lead to frustration of the
inter-plane Josephson coupling, which naturally explains the layer
decoupling seen in 1/8 doped LBCO.\footnote{The problem of the 3D phase transition in a system with an effective layer decoupling is largely unsolved. See, however, the recent work of Raman, Oganesyan and Sondhi.\cite{raman-2009}} In analogy with frustrated magnetic
systems (in which the superconducting order is thought of as an $XY$
pseudo-spin), this frustration can also lead to various forms of
non-collinear order. In the PDW case, such non-collinear orders break
time-reversal symmetry and are accompanied by spontaneous equilibrium
currents.

In this section, we give detailed predictions for the patterns of bulk
time-reversal symmetry breaking and spontaneous currents in various lattice
geometries. We will discuss this problem at zero temperature and at a
classical level. 
It is worth noting that the non-collinear order, where it occurs, results in
a partial lifting of the frustration. In the case of a PDW in the LTT
structure (relevant to {La$_{2-x}$Ba$_{x}$CuO$_{4}$}), we shall show that it
results in a non-vanishing effective Josephson coupling between planes, and
hence, in a sense,
spoils the strict layer decoupling we have touted.
However, this effective Josephson coupling is 
equivalent to a higher order coupling \cite{berg-2007} (due to coherent
tunneling of two Cooper pairs), both in terms of its small magnitude, and
its dependence on the cosine of twice the difference of the superconducting
phases on neighboring planes. (See Eqs. \eqref{E_LTT} and \eqref{Jeff}.) Note
also that defects (such as point defects, domain walls or twin boundaries)
can lead to additional intra-plane time reversal symmetry breaking, that can
drive the system into a glassy superconducting state (as discussed in Sec. \ref{sec:op}).\footnote{An in-plane magnetic field can also change the inter-layer frustration, leading to small violations of the layer decoupling effect. If large enough such effects can be used to detect a PDW state. A similar effect can also take place in junctions between an FFLO state and a uniform superconductor\cite{yang-2000}.}

Let us start with the case of the LBCO LTT\ structure, in which the stripe
direction rotates by $90^{\circ }$ between adjacent planes. We model the
system by a 3D discrete lattice of Josephson junctions, shown in Fig. \ref{fig:spiral}a.\footnote{
Note that we are actually considering a simplified version of the LBCO LTT
structure. The structure in Fig. \ref{fig:spiral}a has two planes per unit
cell, while the LBCO LTT structure has four. The difference is that in LBCO,
the charge stripes in second neighboring planes (which are parallel to each
other) are shifted by half a period relative to one another, while in Fig. \ref{fig:spiral}a they are not. However, the considerations we
discuss here are the same for two structures, and the resulting
non-collinear ground states are similar.} 
The lattice spacing in
the plane is the inter-stripe distance $\lambda$, and $c$ is the
inter-plane distance. Each lattice point has a single degree of
freedom $\theta _{\mathbf{r}}$, which is the local value of the
superconducting phase at that point. $J,-J^{\prime },J^{\prime
\prime }$ are the intra-stripe, the inter-stripe and the
inter-plane Josephson couplings, respectively. We assume that
$J>J^{\prime }\gg J^{\prime \prime }>0$, corresponding to a
unidirectional striped superconductor in the planes. For any
collinear configuration, the Josephson coupling between the planes
vanishes. However, if the staggered order parameter in each plane
is rotated by $90^{\circ }$ relative to its neighbors, then the
energy can be lowered by distorting the phases periodically with
respect to the collinear configuration in each plane. We use a
variational \emph{ansatz} for the phases $\theta_\mathbf{r}$ of
the form
\begin{equation}
\theta_{\mathbf{r}}=\frac{1+(-1)^{z}}{2}y
\pi+\frac{1-(-1)^{z}}{2}\left(x+\frac{1}{2}\right)
\pi+(-1)^{x+y+z}\theta
\end{equation}
where $\mathbf{r}=(x,y,z)$ is the integer valued position vector
($x$ and $y$ are measured in units of $\lambda$, and $z$ is
measured in units of $c$), and the distortion angle $\theta$ is a
variational parameter. The Josephson energy per site as a function
of $\theta$ is
\begin{equation}
E_{\mathrm{LTT}}\left( \theta \right) =-\left( J+J^{\prime }\right) \cos
2\theta -J^{\prime \prime }\sin 2\theta \mbox{\rm .}  
\label{E_LTT}
\end{equation}
The inter-plane coupling energy gain is linear in $\theta $, whereas the
cost in intra-plane coupling energy is quadratic in $\theta $. Thus the
distortion occurs for any non-zero value of the inter-plane coupling $J^{\prime \prime }$. Minimizing Eq. (\ref{E_LTT}), we get
\begin{equation}
\tan 2\theta =\frac{J^{\prime \prime }}{J+J^{\prime }}\mbox{\rm .}
\label{tan2}
\end{equation}
The equilibrium currents across the three types of links are $\mathcal{J}=J\sin 2\theta $, $\mathcal{J}^{\prime }=J^{\prime }\sin 2\theta $ and $\mathcal{J}^{\prime \prime }=J^{\prime \prime }\cos 2\theta =\mathcal{J}+\mathcal{J}^{\prime }$, where Eq. \eqref{tan2} was used in the last
relation. The directions of the currents are as indicated in Fig. \ref{fig:spiral}a. Associated with these currents is a magnetic field with
non-zero components in all three directions. The wavevector associated with
this pattern is $\mathbf{Q}=\left( \frac{\pi }{\lambda },\frac{\pi }{\lambda},\frac{\pi }{c}\right) $, where $\lambda $ is the inter-stripe distance
(for LBCO at $x=1/8$, $\lambda \approx 4a$ where $a$ is the Cu-Cu distance)
and $c$ is the inter-plane distance.

The non-collinear distortion in the above pattern induces an effective
non-zero inter-plane coupling.
However, the effective inter-layer coupling is (taking the limit $J^{\prime\prime}\ll J, J^{\prime}$)
\begin{equation}
J_{\mathrm{eff}}\simeq\frac{(J^{\prime \prime})^2}{4(J+J^\prime)}\mbox{\rm ,}
\label{Jeff}
\end{equation}
and is therefore much smaller than the bare inter-plane coupling $J^{\prime\prime}$.
Note, moreover, that the induced Josephson coupling between two neighboring
planes with PDW superconducting phases $\theta_i$ and $\theta_j$ has the
form $J_{\mathrm{eff}}\cos[2(\theta_i-\theta_j)]$, i.e. its period in the
relative phase is $\pi$.

Next, we consider the case of an orthorhombic structure (such as the LTO
phase of LBCO). In this case, rotational symmetry in the plane is broken in
the same way in every plane, and the stripes are all in the same direction.
However, we assume that due to Coulomb interactions, the charge stripes are
shifted by half a period between adjacent planes. (Such a shift is indeed
observed between second neighbor planes in the LTT\ phase of LBCO, in which
the stripe direction is parallel.) Therefore, the inter-plane coupling is
frustrated due to the resulting \textquotedblleft zigzag\textquotedblright\
geometry. We shall show below that the ground state has spiral order which
partially relieves this frustration. Introducing a spiral twist angle $\theta $, such that $\theta_\mathbf{r}=2x\theta$ (as shown in Fig. \ref{fig:spiral}b), costs an energy $E_{\mathrm{ORT} }\left( \theta \right) $
per stripe, given by
\begin{equation}
\frac{E_{\mathrm{ORT}}\left( \theta \right) }{L}=J^{\prime }\cos 2\theta
-2J^{\prime \prime }\cos \theta
\end{equation}
where $L$ is the length of each stripe. The minimum is for $\cos \theta =
\frac{J^{\prime \prime }}{2J^{\prime }}$. Therefore a spiral distortion
occurs for any $J^{\prime \prime }<2J^{\prime }$. The currents along this
links are $\mathcal{J}^{\prime \prime }=-\mathcal{J}^{\prime }=J^{\prime
\prime }\sin \theta $, and their directions are indicated in Fig. \ref{fig:spiral}b. Each plane carries a uniform current which flows
perpendicular to the stripes, and an equal and opposite current flows
between the planes. The magnetic field associated with these currents is
pointing parallel to the stripe direction, and its lowest Fourier component
is at wavevector $\mathbf{Q}=\left( 0,0,\frac{2\pi }{c}\right)$.

Finally, we turn to the case of a 
domain wall in the PDW order, depicted in Fig. \ref{fig:spiral}c.
(Such a defect is very costly energetically, but it is favored by
a twin boundary in the crystal structure.) The Josephson coupling
across the domain wall vanishes for any collinear configuration.
The energy can be lowered by distorting the phases in the pattern
shown in Fig. \ref{fig:spiral}c, which is closely analogous to the
minimum energy configuration in the LTT case (Fig.
\ref{fig:spiral}a). The superconducting phases $\theta
_{\mathbf{r}}$ are given by
\begin{equation}
\theta _{\mathbf{r}}=\left\{
\begin{array}{c}
(x+\frac{1}{2})\pi -(-1)^y\theta _{x}\text{\ }\mbox{\rm  (}x<
1\mbox{\rm )} \\
 y\pi +(-1)^y\theta _{x}\text{ \ \ \ \ \ \ \ }\mbox{\rm  \
(}x\geq 1\mbox{\rm )}
\end{array}
\right. \mbox{\rm ,}
\end{equation}
where the distortion angle $\theta_{x}$ depends on the distance
from the domain wall $x$. (In our notation, $x=0$ and $1$ are the
two columns on either side of the domain wall.) The energy is
\begin{eqnarray}
\frac{E_{\mathrm{DIS}}\left( \left\{ \theta _{x}\right\} \right)
}{L} &=&-J\sum_{x=1}^{\infty }\cos \left( \theta_{x+1}-\theta
_{x}\right) -J^{\prime }\sum_{x=1}^{\infty }\cos \left( 2\theta
_{x}\right)   \nonumber
\\
&&-J^{\prime }\sum_{x=-\infty }^{0}\cos \left( \theta
_{x-1}-\theta _{x}\right) -J\sum_{x=-\infty }^{0}\cos \left(
2\theta_{x}\right)
\nonumber \\
&&-\tilde{J}\sin \left( \theta_{x=1}+\theta_{x=0}\right) \mbox{\rm
.} \label{E_dis}
\end{eqnarray}
Here, $\tilde{J}$ is the Josephson coupling across the domain
wall, and $L$ is the number of sites along the domain wall. For
simplicity, we assume that $\tilde{J}\ll J,J^{\prime }$, in which
case $\theta_{x}\ll 1$ and we may expand Eq. (\ref{E_dis}) to
second order in $\theta_{x}$. Minimizing $E_{\mathrm{DIS}}\left(
\left\{ \theta _{x}\right\} \right) $, we obtain the following
solution:
\begin{equation}
\theta _{x}=\left\{
\begin{array}{c}
{\theta} _{<}e^{\alpha x}\text{\ \ }\mbox{\rm  \ (}x<1\mbox{\rm )} \\
{\theta}_{>}e^{-\beta x}\text{\ }\mbox{\rm  (}x\geq 1\mbox{\rm )}
\end{array}
\right. \mbox{\rm ,}
\end{equation}
where $\alpha=2\sinh ^{-1}\left( \sqrt{\frac{J^{\prime }}{J}}\right) $, 
$\beta=2\sinh ^{-1}\left( \sqrt{\frac{J}{J^{\prime }}}\right) $, ${\theta}_{<}=\frac{\tilde{J}}{J^{\prime }\left( 1-e^{-\alpha _{<}}\right) +4J}$ and 
${\theta}_{>}=\frac{\tilde{J}}{J\left( 1-e^{-\alpha _{>}}\right)
+4J^{\prime }} $. Associated with the distortion of the
superconducting phases is a periodic pattern of spontaneous
currents, shown in Fig. \ref{fig:spiral}c, with periodicity of two
inter-stripe distances.

Similar considerations apply to an in-plane Josephson junction between a
striped superconductor and a uniform superconductor, if the boundary is
perpendicular to the stripe direction. Therefore, in such a junction time
reversal symmetry is also broken. The critical current is of order $\frac{\tilde{J}^{2}}{\min \left\{ J,J^{\prime }\right\}}$. [This follows from the
same considerations as the effective inter-plane coupling in the LTT case,
Eq. \eqref{Jeff}.] 
It is thus suppressed relative to the critical current of a
Josephson junction between uniform superconductors, which is of
order $\tilde{J}$, as a result of the frustration of the Josephson
coupling across the junction. Similarly to inter-plane coupling in
the LTT case, the period of the coupling between a uniform and a
striped superconductor in the relative phase is $\pi$, \emph{i.e.}
\emph{half} of the period of the coupling between two uniform
superconductors.

\section{Connections and History}
\setcounter{footnote}{0}
\label{sec:history}

The notion of a superconducting state with spontaneously generated
oscillations in the sign of the order parameter has cropped up, under
various guises, a number of times in the past. It is worthwhile to recount
some of these circumstances, not only in the interest of scholarship, but
also to broaden the range of phenomena which can be addressed within the
same conceptual framework.

\subsection{Josephson $\protect\pi$ junctions}

Since the superconducting order parameter is a charge $2e$ scalar
field, it is often assumed that it is possible to think of the
superconducting state as a Bose condensed state of charge $2e$
bosons. In contrast, most classic treatments of the subject
\cite{schrieffer64} emphasize that many features of BCS theory,
especially those associated with quasiparticle coherence factors,
cannot be understood in this way. At the very least, a bosonic
theory is inadequate to capture basic features of the groundstate
of any superconductor which has gapless quasiparticles, either
because of the order parameter symmetry (\emph{e.g.} d-wave) or
because of scattering from magnetic impurities (gapless
superconductor).

Even ignoring the possibility of gapless quasiparticles, there are
qualitative possibilities in a fermionic system that cannot occur
in a bosonic system. A feature of a time reversal invariant
bosonic system is that the ground-state can be chosen to be real
and nodeless. Thus, the order parameter in a Bose-condensed system
must have a phase which is independent of position. The $\pi$
junctions, which we have been discussing, are possible only
because of the composite character of the superconducting order
parameter \cite{spivak-1991}.

There have been several previous theoretical studies which have found
circumstances under which $\pi$ junctions might occur 
\cite{spivak-1991,Rozhkov-1999,lebedev78}. More recently, the
existence of such $\pi$ junctions in the predicted circumstances
have been confirmed by experiment. The first such experiments
\cite{wollman-1993,tsuei-1994}
were significant as the ``phase sensitive'' measurements which definitively
established the d-wave symmetry of the superconducting order in the
cuprates. More recently, however, mesoscopic $\pi$ junctions between two
s-wave superconductors have been constructed and characterized \cite{vandam-2006}. In our opinion, these latter experiments are also landmarks
in the study of superconductivity. They establish that $\pi$ junctions, the
essential ingredient for the existence of striped superconductors, are
physically possible.

\subsection{FFLO states}

In a superconductor with negligible spin-orbit coupling, it is
possible to generate an imbalance in the population of up and down
spin quasiparticles, either by applying a magnetic field in a
geometry in which it predominantly couples to the electron spins,
or by injecting a non-equilibrium population of quasiparticles
from a neighboring ferromagnet \cite{salkola-1998}. In the related
systems of cold fermionic atomic gases, it is possible to vary the
population of up and down spin atoms independently, and to study
the effect of this population imbalance on the superfluid state
\cite{Hulet-2006,Zwierlein-2006,Radzihovsky-2008}. While a first
order quenching of the superconducting state is possible under
these circumstances, there has also been considerable discussion
of the possibility of spatially modulated superconducting states,
so called FFLO states \cite{Fulde-1964,Larkin-1964}. Two distinct
states of this sort have been considered: (1) The FF state
\cite{Fulde-1964}, in which the order parameter has constant
magnitude but a phase which twists as a function of position
according to $\theta = \Delta \mathbf{k}_F \cdot \mathbf{r}$,
where $\Delta \mathbf{k}_F$ is the difference between the up spin
and down spin Fermi momentum. (2) The LO state \cite{Larkin-1964},
in which the order parameter
remains real, but oscillates in sign with a period $L=2\pi/|\Delta \mathbf{k}_F|$.

The LO state is similar in structure to the striped superconductor
considered here. In the order parameter theory presented in Sec.
\ref{sec:landau}, it corresponds to $\lambda_->0$ in Eq.
(\ref{F4}). The parallel with the FF state (which is realized in
the order parameter theory for $\lambda_-<0$) is less crisp, but
when superconducting striped spirals which spontaneously break
time reversal symmetry arise due to the appropriate type of
geometric frustration of the Josephson couplings (as discussed in
Sec. \ref{sec:Tbreaking}), states that are in many ways analogous
to the FF state also occur in striped superconductors. Thus, many
of the physical phenomena we have discussed in this paper are
pertinent to the FFLO phases in more weakly correlated systems,
with the added richness \cite{Radzihovsky-2008} in the case of
cold atomic gasses that there are conserved quantities associated
with the continuous rotational invariance of the underlying
Hamiltonian.

However, the FFLO states arise from the explicit breaking of time reversal
symmetry. Absent a magnetic field, Kramer's theorem implies perfect nesting
between time-reversed pairs of states on opposite sides of the Fermi
surface, so BCS pairing always occurs preferentially at $\mathbf{k}=\mathbf{0}$. This constraint is removed when time reversal symmetry is
explicitly broken. One can think of the FFLO states as taking
advantage of the ``best'' remaining approximate nesting vector,
$\Delta \mathbf{k}_F$, in the two-particle channel. Alternatively,
one can think of the LO state as
consisting of a set of discommensurations \cite{salkola-1998,Radzihovsky-2008} such that the excess spin-up quasiparticles
are incorporated in mid-gap states localized near the core of the
discommensuration.

The energetic considerations that lead to the FFLO states are thus
very different than the strong-coupling physics that gives rise to
the striped superconductor.\footnote{FFLO states  in the absence of magnetic fields have been shown to exist for special band structures in 1D\cite{Datta-2008} and 2D\cite{Kubo-2008}.}
 The fact that the FFLO states
explicitly break time reversal symmetry implies that they are
macroscopically distinct (as phases of matter) from the striped
superconductors that preserve this symmetry. Even in comparison
with striped states which spontaneously break time reversal
symmetry, the distinction remains that the FFLO states have a net
magnetization, while the striped superconductor does not.
Conversely, the FFLO states generally have no particular relation
to other flavors of electronic ordering, while striped
superconductors, as is characteristic of all electronic liquid
crystals, embody a subtle interplay between multiple ordering
tendencies. Specifically, since the striped superconductor seems
to be generally associated with the strong coupling physics of
doped antiferromagnets, there is a natural sense in which
antiferromagnetism, charge density wave formation, and striped
superconductivity are intertwined.

\subsection{Intertwined orders and emergent symmetries}

One explicit way in which the relation between several order
parameters can be more intimate than in a generic theory of
``competing orders'' is if there is an emergent symmetry at low
energies which unifies them. In particular, the order parameter
structure of the PDW state, involving several order parameters
coupled to each other, evokes the $SO(5)$ approach of a unified
description of antiferromagnetism and uniform $d$-wave
superconductivity \cite{Zhang-1997,demler04}. Indeed, by tuning
the parameters of the effective Landau-Ginzburg theory that we
presented in other sections it is possible to achieve an effective
enlarged symmetry which makes it possible ``rotate'' the striped
superconducting order and charge stripe order parameters into each
other. Even if the enlarged symmetry is not exact, a rotation of
the order parameters is possible but with a finite energy cost (
similar to a ``spin flop''.) It is also worth
noting that a symmetry which allows a similar form of unification of 
$d$-wave superconductivity, electron nematicity, and $d$-density wave
order \cite{chakravarty-2001c} (dDW) has recently been found to
exist under special circumstances by Kee \textit{et al.}
\cite{Kee-2008}. It is therefore possible that there could exist
additional forms of striped superconducting states which
interleave these orders.

Thus, it is possible to view the PDW state as a ``liquid
crystalline'' analog of the $SO(5)$ scenario. Indeed, the
possibility of an $SO(5)$ ``spiral'' was discussed previously by
Zhang \cite{Zhang-1998}. However, it should be noted that in the
context of any conventional Landau-Ginzburg treatment of a system
of competing orders, a general theorem \cite{Pryadko-1999}
precludes a sign change of any component of the order parameter,
and hence precludes the existence of spirals. In order to get a
PDW state from an interplay between d-wave superconductivity and
antiferromagnetism, unconventional gradient dependent interactions
between the different order parameters, such as those discussed in
\cite{Pryadko-1999}, must play a significant role in the physics.

In other words, in addition to the standard couplings allowed by a
theory with several order parameters, the existence of a stripe
order (for instance) in the charge order parameter must be able to
induce a texture in the superconducting order as well. A useful
analogy to keep in mind is the McMillan-deGennes theory of the
nematic-smectic transition in classical liquid crystals in which
the nematic order parameter acts as a component of a gauge field
thus coupling to the \emph{phase} of the smectic order, or in blue
phases of liquid crystals. (For a detailed discussion of these
topics in liquid crystals see, \textit{e.g.\/}
\cite{degennes-1993,chaikin-1995}.) In fact, Ref
\cite{Radzihovsky-2008} presents a theory of FFLO states in
ultra-cold atoms with gauge-like couplings (\textit{i.e.\/}
covariant derivative couplings) that relate the stripe (and
spiral) order to the superconducting order.

In addition to the conceptual advantages, noted above, the
liquid-crystal picture of the PDW state offers a direct way to
classify the phase transitions (both quantum and thermal) out of
this state. Thus, in addition to a direct transition to a normal
state, intermediate phases characterized with composite order
parameters, are also possible leading to an interesting phase
diagram. We will explore these issues in a separate publication
\cite{berg-2009}.

\subsection{PDW states in Hubbard and t-J models}

In the context of the cuprates, there have been several studies
looking for a striped superconducting state in the $t-J$ or
Hubbard models. On the one hand, extensive, but not exhaustive
DMRG calculations by White and Scalapino \cite{white-1998a,scalapinoprivate} have
consistently failed to find evidence in support of any sort of
spontaneously occurring $\pi$ junctions. 
On the other hand, a
number of variational Monte Carlo and renormalized mean field
calculations have concluded that the striped superconductor is
either the ground-state of such a model \cite{himeda-2002}, under
appropriate circumstances, or at least close in energy to the true
ground state \cite{raczkowski-2007,capello-2008,yang-2008b}. These
latter calculations are certainly encouraging, in the sense that they
suggest that there is no obvious energetic reason to \emph{rule
out} the existence of spontaneously occurring PDW order in
strongly correlated electronic systems. 
However, the fact remains that  no spontaneous $\pi$-junction formation has 
 yet been observed in DMRG or other ``unbiased'' studies
of the $t-J$ or the purely repulsive Hubbard models, indicating that  there remain 
basic unsettled issues concerning the microscopic origins of $\pi$ junctions.

\section{Final thoughts}
\label{sec:final}

\setcounter{footnote}{0}

In this paper, we have introduced the PDW phase and studied its
properties theoretically. In terms of symmetry, the PDW is distinct from the standard
uniform superconductor. While some
of its properties are similar to those of a uniform superconductor
({\it e.g.}, zero resistance), others are markedly different: most
importantly, the existence of a Fermi surface (and hence a finite
density of states) in the ordered phase \cite{dror,Radzihovsky-2008}, the possibility of
frustration of the inter-layer coupling (depending on the lattice
geometry), and the strong sensitivity to (non-magnetic) disorder.
Generically, the PDW state in the presence of weak disorder is expected to give
way to a ``superconducting glass'' phase, in which the
configuration average of the \emph{local} superconducting order
parameter vanishes, but the Edwards-Anderson order parameter is
non-zero (and hence gauge symmetry is broken).

Even though the ordered PDW state itself is time reversal
invariant, time reversal symmetry breaking is a very natural
consequence of PDW order, either in the superconducting glass phase, or as a
way of relieving the frustration of the Josephson couplings in some crystal structures. 
Specifically, 
frustration can lead to non-collinear ground state configurations
of the superconducting pseudo-spins (representing the local phase of the superconducting order),  which are analogous to the
non-collinear ground states which are often found in frustrated
spin systems. An even more exotic state that can naturally emerge
from a ``parent'' PDW state is a superconductor with a 
charge $4e$ order parameter \cite{berg-2007,berg-2008a,Radzihovsky-2008}, which can result when the CDW part of
the PDW order is melted by either quantum or thermal fluctuations.

The occurrence of PDW states in microscopic models is an
intrinsically strong coupling effect, since PDW order (much
like CDW or SDW) is not an instability of a generic Fermi surface.
In this paper, we have provided a ``proof of principle'' of a 
not-too-contrived, strongly correlated, microscopic model with a PDW
ground state. This model mimics some features of the striped state
found in the cuprates ({\it e.g.}, it has charge stripes separated by
$\pi$-phase-shifted spin stripes). 
Whether a PDW state can be found in more realistic models, which include such features as uniformly repulsive interactions and a d-wave-like order parameter, remains to be settled.

Doped Mott insulators are strongly correlated systems whose ground states have a strong tendency to form liquid-crystalline-like\cite{kivelson-1998} inhomogeneous phases,\cite{emery-1993,kivelson-1993,white-2000,arrigoni-2002,carlson-2004}. 
In this regard, the PDW state is an electronic liquid crystal phase in which  
the superconducting and charge/spin orders do not compete with each other but rather are intertwined. As some of us have noted earlier,\cite{arrigoni-2004,kivelson-2007} the 
observation of a high pairing scale in such an electronically inhomogeneous state 
is suggestive of the existence of an optimal degree of inhomogeneity 
 for superconductivity.  Indeed, recent ARPES data suggest that the stripe order that develops in {\LBCO} does not suppress the pairing scale.\footnote{ARPES data in  {\LBCO} shows a substantial and weakly doping dependent anti-nodal gap accross  $x=1/8$\cite{valla-2006,shen08}, where the signatures of the PDW state are strongest.}  The fact that the pairing scale is large 
 in this material suggests that the development of charge stripe order suppresses the development of superconducting coherence but not pairing. In fact, it  gives credence to the argument that there 
 is a connection between the emergence of charge order and the mechanism of superconducting pairing.\cite{arrigoni-2004,kivelson-2007}

However,  at present, it is unclear to what extent PDW order should be expected to be common where  stripe order occurs.  On the purely theoretical side, PDW order has proven elusive in DMRG studies of models\cite{white-1998a,scalapinoprivate} with entirely repulsive interactions.  Indeed, in a previous publication\cite{berg-2008a}, we showed that in any weakly interacting superconductor, $\pi$ junctions can only occur under exceedingly fine-tuned circumstances.  It is clear from variational calculations\cite{himeda-2002,raczkowski-2007,capello-2008,yang-2008b} that for strong interactions, the differences in energy between the PDW and uniform sign superconducting states in striped systems is relatively small;  what particular features of the microscopic physics tip the balance one way or another is still not clear.  Accordingly, it is not clear, in the absence of unambiguous experimental evidence, whether in the context of the cuprates, we should expect the PDW state to be a rare occurrence, perhaps stabilized by some particular detail of the electronic structure of {\LBCO}, or if instead we should infer that some degree of local PDW order exists in any cuprate in which evidence of local stripe correlations can be adduced.

To close this Section, we turn to discuss the evidence for PDW
states in the cuprate high temperature superconductors. The
analysis of the PDW state was motivated by the experimental
observations on \LBCO.  Having studied the nature of this phase,
we will now discuss to what extent the signatures of the PDW
state are consistent with experiment. Finally, we speculate on the
possible relevance of these ideas to other members of the cuprate
family.

\subsection{Striped SC phases in La$_{2-x}$Ba$_{x}$CuO$_{4}$ and 214 cuprates}

\label{subsec:lbco} 

As already discussed in Sec.~\ref{sec:experiment}, the onset of clearly identifiable 2D superconducting correlations in \LBCO\ with $x=\frac18$ occurs at $\sim40$~K, together with the onset of static spin-stripe order.  It would be natural to associate this behavior with the simultaneous onset of local PDW order; however, an attempt to reach a consistent interpretation of a broad range of results leads to a more nuanced story.

The original motivation for applying the PDW concept to La$_{1.875}$Ba$_{0.125}$CuO$_4$ was to explain the dynamical layer decoupling through the frustration of the interlayer Josephson coupling in the LTT phase \cite{berg-2007,himeda-2002}, as discussed in Sec.~\ref{sec:Tbreaking}. 
It provides a compelling account\footnote{Since {\LSCO} retains the LTO structure to low temperatures, and the spin correlations in the c-direction measured  at zero field are extremely short-ranged \cite{lake-2005}, it is unclear whether the charge stripes in neighboring planes tend to be perpendicular to each other, as in the LTT materials, or parallel but offset by half a period from each other, as in the {\YBCO} bilayers.  In either case, the interlayer Josephson coupling for a PDW would be highly frustrated.}
 for the induced dynamical layer decoupling produced in underdoped {\LSCO} by
a modest $c$-axis magnetic field \cite{basov08}.  
Moreover, the sensitivity of the PDW to disorder 
which limits the growth of the superconducting correlation length within the planes,
provides a natural explanation for the existence of an enormously enhanced
``superconducting fluctuation'' regime, characterized by
enhanced contributions of local superconductivity to the
electrical conductivity and to (strongly anisotropic)
diamagnetism, but with no global phase coherence.  
Thus, it  naturally accounts for the most dramatic aspects of the 
experimental data \cite{li-2007} below the spin ordering temperature $T_{SDW}$.
We consider this strong evidence that the basic ingredients of the theory are applicable to the stripe ordered state of {\LBCO} and closely related materials.
In addition, 
the observed transition at temperature $T_{3D}$ into a state with zero
resistance in all direction 
has a natural interpretation in terms of an assumed PDW state
as the superconducting glass transition \cite{berg-2008a}. Besides having zero
resistance, the glass phase presumably shows no Meissner effect
and zero critical current. If this latter identification is correct, it leads to the further prediction that this phase should be characterized by
various phenomena associated with slow dynamics, characteristic of
spin glasses, as well as with breaking of time reversal symmetry.
The experimental detection of such phenomena below $T_{3D}$ supercurrents) 
 would serve as further confirmation of the existence of a PDW in this material.
(For example,
the glass phase would likely exhibit  a metastable 
zero-field Kerr effect \cite{xia-2007}.)

One can also look for evidence for the PDW in single-particle properties.  One of the key features of the PDW stripes is the gapping of single-particle excitations in the antinodal region, as illustrated in Fig.~\ref{fig:spectrum}; in contrast, the nodal states remain ungapped.  
From the underlying band-structure, one sees that 
the largest contribution to the density of states with energies near $E_F$ comes from the antinodal regions (where the dispersion 
is relatively flat); thus, the onset of local PDW order should have a major impact on 
 properties sensitive to the total density of states.  Conversely, properties that are largely determined by near nodal quasiparticle dynamics, which presumably includes the quasiparticle contribution  to the in-plane conductivity, may be less strongly affected.

Observed striking changes in various transport  properties of several stripe order cuprates can be interpreted in this light as being  suggestive of the appearance of local PDW order at  the onset of {\it charge}-stripe order at $T_{CO}$ (which is generally somewhat higher than $T_{SDW}$).  In \LBCO\ and Nd- and Eu-doped \LSCO, it is observed that the in-plane thermopower drops dramatically below $T_{CO}$ \cite{li-2007,sera-1989,zhou-1997,hucker-1998} as does the Hall resistivity \cite{noda-1999,adachi-2001,takeshita-2004}.  Furthermore, the opening of a superconducting-like gap
as the temperature drops below $ T_{CO}$ results in an observed \cite{homes-2006} suppression of the in-plane   optical conductivity at frequencies below 40~meV.  In contrast, the  in-plane DC-resistivity changes relatively little \cite{li-2007,ichikawa00} upon cooling through $T_{CO}$.

Putting aside the issue of the onset-temperature, the notion that stripe ordered cuprates exhibit local PDW order  is also supported by ARPES studies.  For example, measurements on
stripe-ordered La$_{1.48}$Nd$_{0.4}$Sr$_{0.12}$CuO$_4$ at $T=15$~K ($> 2T_c$) reveal a
gapless nodal arc of states covering roughly a third of the
nominal Fermi surface, as well as a gap reaching 30 meV in the antinodal
region \cite{chang-2008b}.  Temperature-dependent ARPES measurements on  \LBCO\ with $x=\frac18$ indicate that, for temperatures above the spin-ordering transition, there
is a gapless nodal arc of states, together with a substantial
antinodal gap \cite{shen08}. 

However, there are several aspects of this story which require further analysis.  Firstly, there is the issue that different aspects of the crossovers we would like to identify with the onset of local PDW order appear to onset at different temperatures.  This is not necessarily inconsistent, as a crossover (as opposed to a phase transition) can appear to occur at somewhat different temperatures depending on what quantity is measured and how the data is analyzed.  Nonetheless, the drop in the thermopower and Hall number appears to have a very sharp onset at $T_{CO}$, while the superconducting like drop in the in-plane resistivity at $T_{SDW}$ is also very sharp, at least in 1/8 doped {\LBCO}.  [In this sense, it is reminiscent of the situation \cite{khaykovich-2002} in O doped {\LCO}, where the sharply defined spin ordering and superconducting ordering transitions occur at the same temperature (in zero field) with very small uncertainty.]

A still more perplexing issue arises in correlating the 
onset of the signatures of 2D superconductivity in 
La$_{1.875}$Ba$_{0.125}$CuO$_4$ with the thermal evolution of the ARPES \cite{valla-2006,shen08} spectrum.  Below $T_{SDW}$,
there is clear evidence of the appearance of a $d$-wave-like gap in the nodal region,
with the scale of this second gap being smaller than the pre-existing antinodal gap \cite{shen08}.   This behavior suggests that uniform 
$d$-wave superconductivity develops and coexists with the PDW
superconductor below $T_{SDW}$.  However, this is somewhat problematic, as
the proposed explanation of the dynamical interlayer decoupling and the bounded growth of superconducting correlations that occurs below $T_{SDW}$ rests on the assumed  (near) absence of a uniform component of the order parameter in each plane.
Reconciling the uniform $d$-wave component of the order parameter inferred spectroscopically from ARPES studies of {\LBCO} with the 
apparently almost complete absence of such a component inferred from bulk transport measurements
on the same material is a challenge for future work.
It may be significant, however, that ARPES studies  of 
{\LNSCO} \cite{chang-2008b} and
La$_{1.8-x}$Eu$_{0.2}$Sr$_x$CuO$_4$ \cite{zabolotnyy-2008} appear consistent with pure PDW order, ({\it i.e.}, there is no d-wave gap in the nodal region), although the PDW appears to
set in at around $T_{CO}$, which can be substantially greater than $T_{SDW}$ in
these materials.

\subsection{Dynamical layer decoupling and quasi-two-dimensional behavior in the cuprates}

The cuprate superconductors are layered materials with varying degrees of quasi-two-dimensional behavior. Evidence for quasi-2D behavior (and for dimensional crossover) in the cuprates has existed for a long time and it is well documented. It is thus useful to compare and contrast this well known behavior with the unexpected layer decoupling effect observed in  {\LBCO}.

In a quasi-2D system, as a continuous thermodynamic superconducting phase transition is approached, the in-plane correlation length grows very rapidly. While at first the fluctuations have a markedly 2D character, very close to the phase transition they rapidly cross over to their ultimate 3D behavior. 
Dimensional crossover is 
observed,
for instance,
 in dynamical probes of 
some cuprates. High frequency ($\sim$100 GHz) conductivity measurements in {\BSCCO} (the most quasi-2D material among the cuprates)  by 
Corson et al \cite{Corson-1999} showed that (at those frequencies) the fluctuation conductivity is 2D-like and exhibits
Kosterlitz-Thouless behavior, as if the CuO$_2$ planes were effectively decoupled.  
Similarly, quasi-2D behavior in the dynamic conductivity (with frequencies in the range 1-10 GHz) has been observed in underdoped {\LSCO} near $T_c$ (but not in overdoped {\LSCO}) by Kitano et al \cite{Kitano2006}. 
By probing the system at finite frequency, these experiments explore the correlations at a frequency dependent mesoscopic length scale, where sufficiently weak 3D couplings have negligible effect on the physics.
By contrast, 
the resistive transition both in {\BSCCO} and in {\LSCO}, measured at zero frequency in macroscopic samples, is not of the 2D XY (Kosterlitz-Thouless) type, but rather reflects the three-dimensional 
nature of these materials. 

In contrast, the unusual layer decoupling effect observed in {\em stripe-ordered} {\LBCO} takes place in a temperature range  where the CuO$_2$ planes appear to become superconducting (well above the three-dimensional critical temperature).\cite{li-2007,tranquada08} The layer decoupling effect is observed in the resistive transition.
and is thus not a dimensional crossover effect. As we noted above, in this regime {\LBCO}  behaves as if for some reason the effective inter-layer Josephson coupling is either turned off (which is unphysical) or 
 is somehow {\em frustrated}.

Support for this idea is provided by recent Josephson resonance experiments in {\LSCO} by Schafgans et al\cite{basov08}, which essentially measure
the c-axis superfluid stiffness, $\rho_c$. 
In the absence of an external magnetic field, $\rho_c$ 
has the expected\cite{basov-1994,dordevic-2002,homes-2004} magnitude, {\it i.e.} $\rho_c$ is proportional to the normal state conductivity at $T_c$.
However, for underdoped materials, $\rho_c$ becomes unmeasurably small
in the presence of moderate magnetic fields ($ B \leq 8T$).  Magnetic fields  are known to induce
static spin-stripe order (as detected by neutron scattering experiments\cite{lake-2002}) in precisely the same range of field strengths and hole concentration. These experiments thus suggest\cite{basov08} that the ``fluctuating stripe order'' \cite{kivelson-2003}
seen in {\LSCO} at zero field may actually be of the PDW type
and that dynamical layer decoupling occurs as static stripe order is stabilized in a magnetic field.\footnote{While it is tempting to reinterpret in hindsight the results of Corson et al\cite{Corson-1999} as being indicative of ``fluctuating PDW order'' in {\BSCCO} we should note that the STM data on this material\cite{kohsaka-2007} show a glassy pattern of short range stripe order at high bias. However, as we explained elsewhere in this paper, a glassy version of the PDW state would not exhibit a sharp layer decoupling effect.}
Indeed, in materials, including {\LBCO} and {\LNSCO},  which exhibit stripe order in zero field, $\rho_c$ is found\cite{Schafgans-2009} to be orders of magnitude smaller than its ``expected'' value on the basis of the normal state conductivity.

\subsection{Possible relevance to other cuprates}

Although there are still open issues, the PDW state (or its glassy version) seems to 
offer a rather compelling explanation for what is otherwise an extremely surprising set of phenomena observed in stripe ordered cuprates.   Could these ideas also be relevant to a broader range of phenomena in the cuprates? The direct empirical information available \cite{kivelson-2003} concerning the structure of any sort of static or fluctuating stripe order present in cuprates outside the 214 family is much less clear.\footnote{The results of recent neutron scattering studies of underdoped {\YBCO} by Hinkov {\it et al.} \cite{hinkov-2007b} have confirmed \cite{ando-2002} the existence of a nematic phase, derived from the weak melting of a stripe ordered  state, onsetting below a temperature comparable to the  pseudogap onset-temperature, $T^\star$.  Still more recently \cite{hinkov-2008}, the same authors have demonstrated that modest magnetic fields stabilize static spin-stripe order where primarily fluctuating (nematic) order existed at zero field.}
Consequently, any attempt to achieve a theoretical understanding based on the assumed existence of a PDW state is necessarily speculative.  We therefore present the discussion of this final section in the spirit of provocative conjectures, which we believe are deserving of further investigation.

ARPES studies of \BSCCO\ \cite{Kanigel} and \LSCO\ \cite{shi-2008,yoshida-2008} have revealed ``Fermi arcs'' of gapless states between antinodal pseudogaps.  There has been a great deal of controversy over the nature of the antinodal pseudogap \cite{norman-2005,hufner-2008}.   Two recent studies of \BSCCO\ have reported signatures of
Bogoliubov quasiparticles in the antinodal gap region \cite{yang08,kanigel-2008}, which was interpreted as being suggestive that the pseudogap is, at least in part, produced by superconducting fluctuations.
On cooling 
through $T_c$, a $d$-wave gap appears along the nodal arc \cite{shi-2008,yoshida-2008,lee-2007}.  In near optimally doped samples, as $T\to 0$, this d-wave gap and the pseudo-gap merge to form a single gap with a simple $[\cos(k_x)-\cos(k_y)]$ form.  However, in underdoped samples, even as $T\to 0$, the nodal gap appears to have a different energy scale than the antinodal gap ({\it i.e.}, they do not merge to form a simple $d$-wave gap) \cite{yoshida-2008,lee-2007}.
Thus, in some ways it is clear that there are two distinct gaps - an antinodal pseudo-gap that might be associated with some sort of ``competing'' order, and a nodal gap, which is clearly superconducting in the sense that it onsets quite sharply at $T_c$.  However, in other ways it seems that all the gaps have some unifying superconducting character.

We propose that this puzzle may be resolved by postulating that there are two distinct gaps, both with superconducting character in the sense that one is associated with uniform  the other with modulated superconducting order.  Indeed,  the measured quasiparticle spectral function in the pseudogap 
 looks somewhat like that of the PDW state (see Fig. \ref{fig:spectrum}).\footnote{Technically, the electron-hole-mixed quasiparticles in the antinodal region of a PDW state are not perfectly symmetric with respect to $E_F$ (see Fig.~\ref{fig:spectrum}), in contrast to Bogoliubov quasiparticles; however, to detect this distinction, one would need to measure a sample containing a single-domain PDW state.  Measurements on a nematic PDW state in {\BSCCO} would average over stripe orientations; furthermore, the experimental ``quasiparticle'' peaks are quite broad in the pseudogap state \cite{yang08,kanigel-2008}, so that any fine details are hidden by damping.  The overdamping also fills in the spectral weight at $E_F$, in contrast to the true gap that is found for $T<T_c$ \cite{yang08,shi-2008}.} 
Moreover, just such a combination of modulated and uniform superconducting orders has been previously proposed on phenomenological grounds  to explain \cite{podolsky,robertson} STM spectra \cite{howald-2003a,howald-2003b,hoffman-2002b,gomes-2007,slezak-2008} in {\BSCCO} and other cuprates \cite{kohsaka-2007,wise-2008}.

Seemingly more direct evidence of superconducting fluctuations in the normal state of \LSCO, \BSCCO, and Bi$_2$Sr$_{2-y}$La$_y$CuO$_6$ has been reported by Ong and coworkers \cite{wang-2002a,ong-2007} based on measurements of the Nernst effect and diamagnetism.   Nernst measurements on \YBCO\ \cite{rullier-2006} and STM studies of \BSCCO\ \cite{pasupathy-2008} suggest that disorder may be important to the existence of the fluctuation effects over a substantial temperature range.  It is intriguing that the onset temperatures of the enhanced Nernst response in \LSCO\ has a maximum at $x\sim0.1$ \cite{wang-2002a}, close to the optimum doping for stripe order.  
Moreover, Taillefer and coworkers \cite{Taillefer-2008} have found close correlations between an enhanced Nernst signal and stripe order. 
 Neither the observed sensitivity to disorder nor the association with stripe order, by themselves, necessarily negate the interpretation of these effects in terms of superconducting fluctuations; however, both would be unusual in the case of a simple, homogeneous d-wave superconductor.  
While we are far from having an explicit theory, it seems to us that these general trends are consistent with the existence of a disordered PDW state over at least a portion of the pseudogap phase.   
Specifically, Ong and coworkers \cite{li-2005} have reported the observation of a sublinear dependence of the magnetization on magnetic field ($M \sim -B^{\alpha}$ with $\alpha < 1$) in a relatively narrow  but non-vanishing range of temperatures above $T_c$ in crystals of {\BSCCO}.  This behavior, if it truly persists in the limit $B \to 0$, must signify the existence of a distinct phase of matter in this range of temperatures, which we very tentatively propose could be a superconducting glass formed from a disordered PDW.\footnote{Some of us have argued elsewhere \cite{emery-1997,kivelson-2007} that the pair correlations in hole-rich stripes correspond to spin singlet correlations, so that the pairing energy is reflected in the singlet-triplet excitation energy.  The description of pairing and spin correlations within the charge stripes has much in common with the RVB perspective \cite{anderson-2004}; however, electronic self-organization into stripes certainly enhances, and may be necessary to realize this behavior in the CuO$_2$ planes \cite{emery-1997,kivelson-2007}.}  

One of the most intriguing recent discoveries in the cuprates involve several distinct observations of  a rather subtle, and not fully understood, form of time reversal symmetry breaking in the pseudogap phase of {\YBCO} \cite{fauque-2006,xia-2007} and HgBa$_2$CuO$_{4+\delta}$ \cite{greven-2008}.  As we have seen, various forms of subtle time-reversal symmetry breaking can occur when frustration is added into the PDW mix.  It is our hope that, with further work, a relation can be established between these two rather vague statements.

\ack
We thank Peter Abbamonte, Dimitri Basov, Hong Yao, Ruihua He, Srinivas Raghu, Aharon Kapitulnik, Eun-Ah Kim, Vadim Oganesyan,
Gil Refael, Doug Scalapino, Dale Van Harlingen, Kun Yang, and Shoucheng Zhang  for great discussions. This work was
supported in part by the National Science Foundation, under grants DMR
0758462 (E.F.) and DMR 0531196 (S.A.K.), and by the Office of Science, U.S.
Department of Energy under Contracts DE-FG02-91ER45439 through the Frederick
Seitz Materials Research Laboratory at the University of Illinois (E.F.),
DE-FG02-06ER46287 through the Geballe Laboratory of Advanced Materials at
Stanford University (S.A.K. and E.B.), and DE-AC02-98CH10886 at Brookhaven
(J.M.T.).

\section*{References}

\providecommand{\newblock}{}

\end{document}